\documentclass{emulateapj}
\usepackage{natbib}
\usepackage{graphicx}
\usepackage{mathrsfs}
\usepackage{amsmath}
\usepackage{tikz}
\usepackage{hyperref}

\tikzstyle{blue} = [rectangle, rounded corners, minimum width=3cm, minimum height=1cm,text centered, draw=black, fill=blue!30]
\tikzstyle{red} = [rectangle, rounded corners, minimum width=3cm, minimum height=1cm,text centered, draw=black, fill=red!30]
\tikzstyle{green} = [rectangle, rounded corners, minimum width=1.5cm, minimum height=1cm,text centered, draw=black, fill=green!30]
\tikzstyle{orange} = [rectangle, rounded corners, minimum width=1.5cm, minimum height=1cm,text centered, draw=black, fill=orange!30]
\tikzstyle{arrow} = [thick,->,>=stealth]

\begin{document}

\title{Constraints on Evolutionary Timescales for M Dwarf Planets from Dynamical Stability Arguments\vspace{-3em}}
\author{Katie Teixeira}
\affiliation{Department of Astronomy, University of Texas at Austin, Austin, TX 78712, USA}
\author{Sarah Ballard}
\affiliation{Department of Astronomy, University of Florida, Gainesville, FL 32611, USA}

\begin{abstract}
The diversity of dynamical conditions among exoplanets is now well established. Yet, the relevance of orbital dynamical timescales to biological evolutionary timescales is poorly understood. Given that even minor orbital changes may place significant pressure on any organisms living on a planet, dynamical sculpting has important implications for the putative evolution of life. In this manuscript, we employ a Monte Carlo framework to investigate how a range of exoplanetary dynamical sculpting timescales affects timescales for biological evolution. We proceed with minimal assumptions for how dynamical sculpting proceeds and the emergence and persistence of life. We focus our investigation on M dwarf stars, the most common exoplanetary hosts in the Milky Way. We assign dynamical statuses, dependent on stellar age, to a suite of planetary systems, varying the rate of dynamical disruption within limits that are consistent with present-day planet demographics. We then simulate the observed yield of planets according to the completeness of NASA’s \textit{Kepler} and \textit{TESS} missions, and investigate the properties of these samples. With this simplified approach, we find that systems hosting multiple transiting planets ought to have, on average, shorter dynamically-uninterrupted intervals than single-transiting systems. However, depending upon the rate of dynamical sculpting, planets orbiting older stars will exhibit the opposite trend. Even modest constraints on stellar age would help identify ``older" stars for which this holds. The degree of these effects varies, dependent upon both the intrinsic dynamical demographics of exoplanets and whether we consider planets detected by NASA's \textit{Kepler} or \textit{TESS} missions.
\end{abstract}

\keywords{M dwarfs, transits, habitability}

\section{Introduction}
With thousands of confirmed exoplanets now in hand, questions about the search for life elsewhere have shifted from {\it whether} potential sites exist to {\it which} of the multitude to prioritize. More than 5000 exoplanets have been discovered and confirmed with NASA's {\it Kepler} space telescope and \textit{Transiting Exoplanet Survey Satellite} $\left({\it TESS}\right)$ \citep{Borucki09,Ricker14}. These large-scale transit surveys have enabled tremendous progress toward understanding the sizes and orbital properties of exoplanets. Of particular interest is the frequency of rocky planets residing in their host stars’ “habitable zones.” Beyond the identification of new planets, the {\it James Webb Space Telescope} $\left({\it JWST}\right)$ enables the characterization of their atmospheres, giving astrobiologists a much clearer picture of their potential habitability. However, observing time with {\it JWST} is an extremely limited and precious resource. The selection of exoplanetary systems for atmospheric follow-up study, prioritizing those most likely to harbor life, will be a critical endeavour.

In particular, planets orbiting M dwarfs have emerged as likely targets for detailed follow-up study. Firstly, they are numerous: 70 percent of stars in our galaxy are M dwarfs, \citep{Henry04}. Secondly, they often host small planets: on average, there are $\ge$2.5 planets with radii between 1 and 4 Earth radii orbiting M dwarfs, per \cite{Dressing15}. Thirdly, planets, and correspondingly their atmospheres, produce more detectable signatures around smaller stars due to larger transit depth \citep{Tarter07, Shields16}. For example, 200 hours of {\it JWST} time is sufficient to extract a high signal-to-noise detection of biomarkers like oxygen on an M dwarf planet, while the same planet orbiting an FGK dwarf would require orders of magnitude more observational time \citep{Kaltenegger09}.

Yet, while M dwarf planetary systems are ubiquitous in the Milky Way, they are not drawn from a single blueprint. Many host compact, dynamically cool (i.e. low eccentricity and mutual inclination) systems of terrestrial planets, with the TRAPPIST-1 system typifying this category \citep{Gillon17}. Yet, every M dwarf hosting a TRAPPIST-like system would produce planetary yields incompatible with the findings of \textit{Kepler}: per \cite{Muirhead15}, only 20 percent of mid-M dwarf stars host ``compact multiple" systems, systems with multiple planets that orbit with periods less than 10 days. In fitting the {\it Kepler} M dwarf multi-planet yield, \cite{Ballard16} estimated an average of 6 planets per system and average mutual inclination of 2 degrees among 20\% of M dwarfs, with the remaining 80\% likely hosting less planets or planets in dynamically hotter configurations (i.e. high eccentricity and mutual inclination). The origin for this diversity among M dwarf systems is as-yet only partly understood. It could be attributable to formation conditions alone \citep{Dawson16, Moriarty16, MacDonald20}. The link between the ``compact multiple'' occurrence rate around M dwarfs to stellar metallicity \citep{Anderson20} favors this hypothesis. Alternatively, ongoing dynamical sculpting on timescales up to Gyr may also produce the observed mixture of dynamical temperature among M dwarf systems \citep{Pu15}, whether it occurs by self-excitation or by larger perturbing companions \citep{Becker17}. It is this latter hypothesis that we explore here: given the assumption that planetary systems are metastable at birth, we aim to explore how a range of metastability timescales translates to putative evolutionary timescales.

The number of conditions that inform whether life evolves on the surface of a planet, and the degree to which each matters, is presently deeply uncertain (for a summary of the planetary properties that \textit{may} matter, see \citealt{Kopparapu19}). It is useful to define a timescale over which biological evolution may proceed, if conditions are met, possibly leading to the emergence of complex multi-cellular life \citep{Dong_19,Knoll_15}. We aim here to establish only an upper bound on the number of planets hosting life with evolutionary timescale $\tau$, where $\tau$ can lie between zero (for a planetary system that just formed) and the age of the Milky Way (for a planetary system that formed early in the life of the galaxy). Though life \textit{may or may not} evolve on the surface of a given planet, we can state with certainty that evolution has \textit{not} proceeded there for longer than the age of its star. For the sake of this experiment, we define ``evolutionary timescale" to be the duration of time that a planet has existed in a dynamically quiescent state. If dynamical excitation occurs, and planets collide, we assume that complete mass extinction occurs and that the ``evolutionary clock" is reset to zero. That is, if life \textit{has} evolved, evolution cannot proceed if the surface has been rendered molten by a recent collision. If dynamical excitation has instead occurred by some process other than collision, say migration, we assume that evolution may proceed; while a large change in eccentricity, for example, may induce mass extinction (resetting the ``clock" to zero) of some species, some other life forms may be robust to it. 

We organize this manuscript as follows. In Section \ref{sec:Methods}, we describe our simulated sample of M dwarf planets. We detail how we vary our prescription for the way that dynamical sculpting proceeds, and how we correspondingly assign a dynamical stability status to each planetary system. We describe how the properties and evolutionary timescales $\tau$ of planets are assigned, and how we ``observe'' the transits of our synthetic sample of planets using both {\it Kepler} and {\it TESS} approximations for survey completeness. In Section \ref{sec:Results}, we analyze large-scale demographics of our inherent and observed samples and the resulting distributions of $\tau$ among our samples. In Section \ref{sec:Discussion}, we investigate whether transit multiplicity is an informative metric for $\tau$: that is, whether transit multiplicity is predictive for how long a system has been in a quiescent dynamical state. In Section \ref{sec:Conclusions} we summarize our findings and conclude. 

\section{Methods}
\label{sec:Methods}
If large-scale dynamical sculpting is operative in M dwarf planetary systems, it must be consistent with their known demographic properties. Namely, any proposed sculpting law, when applied to a suite of synthetic planetary systems, must first replicate the observed bulk properties, specifically the dynamical properties, of real M dwarf systems. Beyond the properties of any one planet, both the ``dynamical temperature" \citep{Tremaine12, Tremaine15} and the ``angular momentum deficit" (AMD; \citealt{Laskar17}) quantify the dynamical status of the system as a whole. This ``deficit" encodes the departure of the system from its state of maximum possible angular momentum, which occurs when the mutual inclination and eccentricity of all planets are equal to zero. When the difference between the maximum state and the actual state is large (``high" AMD), inclinations and eccentricities are higher. A ``low" AMD means that low eccentricities and mutual inclinations place the system's total angular momentum near its maximum theoretical value. 

Specifically, the AMD of an entire system is represented as the sum of the AMD of its $j$ individual planets, so that 
\begin{equation}
    \mathrm{AMD}=\sum_{0}^{j}\mathrm{AMD}_{j},
    \label{eq:AMD}
\end{equation}
where the AMD of individual planets depends upon the mass of the planet $M_{p}$, the mass of the star $M_{\star}$, the semi-major axis $a$, the eccentricity $e$, and the orbital inclination $i$:
\begin{equation}
    \mathrm{AMD}_{j} = M_{p,j} \sqrt{GM_{\star}a_{j}} \left(1 - \sqrt{1 - e_{j}^2} \cos(i_{j})\right)
\label{eq:AMD_2}
\end{equation}

\cite{He20} applied the AMD framework to interpret the range in dynamical excitation among planetary systems, a phenomenon that has been modeled as both a unimodal distribution in $\{e,i\}$ space (as in that work, as well as \cite{Zhu18}), and as a multimodal distribution (e.g. \citealt{Xie16}, \citealt{Vaneylen14}, and \citealt{mills_california-kepler_2019}). More specifically, \cite{He20} demonstrated that the distribution of $\{e,i\}$ consistent with observed planet statistics such as multiplicity, period ratio, and transit duration ratio, match that of a sample in which all multi-planet systems are at the maximal AMD ``stability limit" formulated by \cite{Laskar17}. This is based on the hypothesis that a series of collisional events during planet formation will decrease total system AMD, so that systems evolve from having high AMD, unstable orbits to having a total AMD just below a critical value under which they have stable orbits. In this work, we will employ observationally-derived distributions in $\{e,i\}$ for M dwarfs, as well as number of planets per system and their spacing. We will adopt a multimodal distribution to model our planetary systems, consistent with our hypothesis of dynamical sculpting, whereby planetary systems ``start" dynamically cool (low AMD) and evolve to be dynamically hotter (high AMD). By tracking AMD through this process, we aim (1) to show that the multimodal distribution ultimately shows broad consistency with the theoretical AMD stability limited distribution of $\{e,i\}$, and (2) to model the theoretical AMD evolution of the ensemble of planetary systems over long timescales.

\subsection{Generation of Planetary Sample}
\label{sec:sample}
We simulate a large number of M dwarf planetary systems ($N_\mathrm{tot} \sim 10^6$), subdivided into \textit{Kepler} and \textit{TESS} mission-sized samples. We assign system properties and then individual planet properties. We then simulate observations of the planetary system samples by \textit{Kepler} and \textit{TESS}. We note that we use a synthetic sample as opposed to the set of M dwarfs observed by \textit{Kepler} and \textit{TESS} because we are attempting to study a possible population-level phenomena, which requires the exact properties of M dwarfs to be known.

\subsection{System Properties}
\label{sec:System-Properties}

\begin{figure}
    \centering
    \includegraphics[width=3.4in]{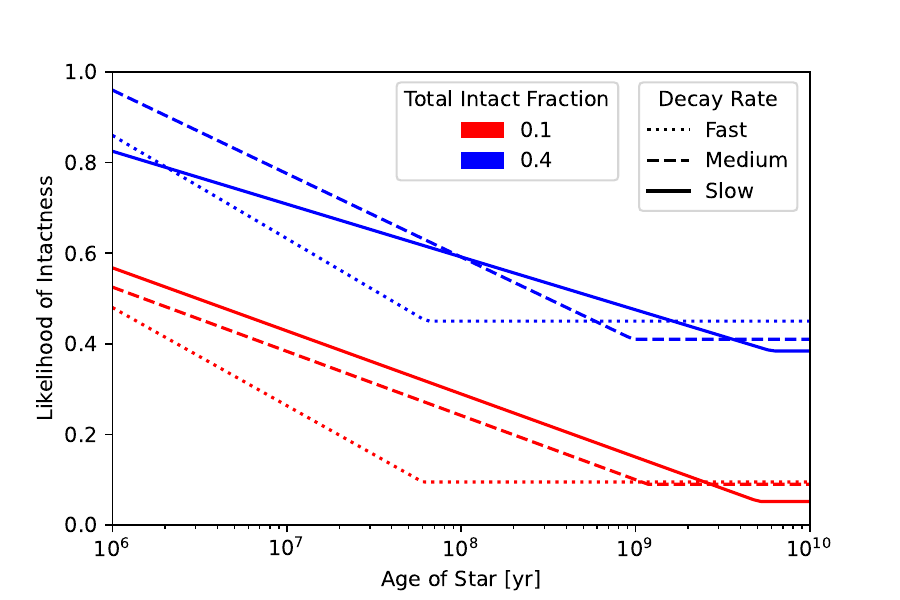}
    \caption{Functions of the likelihood of intactness with respect to stellar age for different combinations of total intact fraction and decay rate.}
    \label{fig:LoI}
\end{figure}

We initialize each system with an M dwarf star of mass $M_{\star}=0.5M_{\odot}$ and radius $R_{\star}=0.5R_{\odot}$. While both \textit{Kepler} and \textit{TESS} missions observed a range of M dwarf spectral types, we adopt a simplying assumption of host star mass and radius. We justify this choice by noting that errors on \textit{Kepler} and \textit{TESS} M dwarf measured masses and radii tend to be between 10-20\%, and our exact chosen values typically fall within those error bars. It is important to note that changing stellar radius does affect transit probability (i.e., a transit is more likely around a larger star). A radius of $0.5R_{\odot}$ is average for \textit{Kepler} and conservative for \textit{TESS}, so our simulations would tend to overestimate total transit yields for \textit{TESS}. However, our goal in this paper is not to quantify the \textit{TESS} yield but to compare two populations of M dwarf planetary systems. The ability to detect planets in these different populations should increase or decrease at similar rates based on the radius of the host star. Therefore, while we understand that the \textit{Kepler} and \textit{TESS} samples are more complicated than we model here, our assumptions do not dramatically affect the results we present.

We must then assign each star an age based on an age distribution of M dwarfs in the Milky Way. Unfortunately, it is notoriously difficult to determine the ages of M dwarfs with any accuracy, as they evolve over very long timescales (see e.g. \cite{Shields16} for a summary). Assuming that the star formation rate is constant in the Milky Way from its formation to present day \citep{Feiden_2021}, we draw stellar age $A$ from a uniform distribution,

\begin{equation}
\label{eq:System-Age}
    A \sim \mathscr{U}(10^{6}, 10^{10}\ \mathrm{yr}).
\end{equation}

We now consider the functional form of the hypothetical disruption mechanism. Dynamical instabilities can manifest over timescales that span many orders of magnitude, so the rate at which systems move from dynamically cold to dynamically hot must depend upon the mechanism driving instability. In systems of three or more planets on initially circular orbits, there is a minimum semimajor axis spacing below which three-body mean motion resonances (MMRs) overlap and drive chaos \citep{Quillen11, Petit20}. \cite{Petit20} show that such instabilities occur within $\sim 10^8$ orbits for a wide range of planetary masses, which in observed systems corresponds to only a few Myr. While systems with sufficient spacing between adjacent planets can typically remain stable over long timescales \citep{SmithLissauer09}, there is a critical eccentricity above which two-body mean-motion resonances can overlap and drive instabilities for more widely separated pairs of planets \citep{Hadden18}. While the timescales for such instabilities are not yet fully understood, N-body integrations find that they typically occur on timescales $\lesssim 10^9$ orbits \citep{Tamayo20, Tamayo21}, or a few tens of Myrs for typical systems. In summary, instabilities driven either by two-body or three-body MMRs would occur early on in systems' lifetimes. In contrast, both perturbations due to a distant companion and secular chaos would be operative on timescales closer to the Gyr lifetimes of the planetary systems. The von Zeipel-Lidov-Kozai effect \citep{Lidov62, Kozai62, Naoz11, Lithwick11EKL, Naoz16, Takashi19} can drive instability on longer timescales, though for planets in compact multiple configurations (common among M dwarfs) the interactions between adjacent planets are more important \citep{Innanen97}. \cite{Liu15}, \cite{Denham19}, and \cite {Wei21} investigated the conditions that modulate which effects are dominant. Secular chaos can also cause instability on longer timescales \citep{Laskar00, Lithwick11, Laskar17, Petit17}. We assume that when dynamical disruption occurs, it is relatively quick: that is, any intermediate state between dynamically cold and dynamically hot is much shorter than the lifetime of the star, so that we can approximate the transition as immediate, and call it a ``disruption''. Regardless of the mechanism, we assume that these ``disrupted'', dynamically hot systems possess higher eccentricities and relative mutual inclinations; this also means that in these systems, there are fewer planets and those planets have wider spacings.

While there are many different effects that could contribute to dynamical instabilities in planetary systems, we choose to base the functional form of our disruption mechanism on the work of \cite{SmithLissauer09}. \cite{SmithLissauer09} used numerical simulations to model Earth-sized planets on initially circular, evenly-spaced orbits and found that systems decayed based on a piecewise function depending on the spacing between planets. Specifically, at spacings lower than some critical value, systems became unstable at $\sim$10 years, while at intermediate spacing, the base-10 logarithm of the stability timescale grew linearly with spacing. Finally, at sufficiently high spacings, \cite{SmithLissauer09} found an upturn in the base-10 logarithm of stability time with spacing, such that these systems would be indefinitely stable. It is plausible that decay of this form could transform a population of only dynamically cold systems into a mixture of dynamically hot and cold systems. \cite{Pu15} further explored this concept as applied to the \textit{Kepler} dataset, in an attempt to attribute the \textit{Kepler} dichotomy to a before-and-after scattering event. \cite{Pu15} formulated a decay function of the same form as in \cite{SmithLissauer09} that match present-day Kepler demographics, noting that the exact parameters of the decay function are not well known.

Without making an assumption about the parameters of the decay function, we employ a range of hypothetical timescales on which systems could disrupt. Based on a prescription for dynamical sculpting of exoplanets, the system is assigned a boolean dynamical state $S$ -- ``intact" ($I$) or ``disrupted" ($D$) -- according to its age.  The decay rate $\mathscr{D}$ is the rate at which systems become disrupted in a sample as the system age increases. The total intact fraction $\bar{\mathscr{L}}$ is the fraction of systems which are intact from a representative sample of systems with uniformly distributed ages. We adopt some simplifying language here, with respect to the dynamical instability timescales. Systems could become disrupted very early in their lifetimes, with sculpting ceasing before 100 Myr (``fast" decay). Sculpting could also be operative later in the system lifetime, ceasing at 1 Gyr (``medium" decay), or on the even longer timescale of 5 Gyr (``slow" decay). While there exists theoretical support for dynamical disruption on timescales of years \citep{Pu15} to Gyr \citep{Batygin09}, we require only consistency with the observed ``intact rate'' among M dwarf planetary systems today. \cite{Muirhead15} estimated this value to be 0.2, but with a 1$\sigma$ confidence interval ranging between fractions of 0.1 and 0.4 \citep{Ballard19}. 

We therefore test nine distinct fiducial dynamical instability laws based on the form found in \cite{SmithLissauer09} and explored in \cite{Pu15}. We select a ``slow", ``medium", and ``fast" sculpting law for each of three resultant modern ``intact" fractions: 0.1, 0.2, and 0.4. We note that our 0.2, ``fast" decay function closely resembles that in \cite{Pu15}, while our family of functions generally brackets other possibilities that are supported in the literature. In this way, we adopt an agnostic approach to the mechanism driving instability, mandating only that it result in an intact fraction today that is within 1$\sigma$ of its observed value.  

We define a likelihood of intactness (LoI) function of age to calculate the likelihood $\mathscr{L}(A)$ that a system with age $A$ is intact. A random number $r$ between 0 and 1 is chosen, and the system is assigned intact if $r$ is less than $\mathscr{L}(A)$, otherwise it is assigned disrupted:
\begin{equation}
    S=
    \begin{cases}
        I & r<\mathscr{L}(A)\\
        D & r>\mathscr{L}(A)\\
    \end{cases}
\end{equation}
Of our 9 different LoI functions, 6 are shown in Figure \ref{fig:LoI}. Each LoI function is of the following form,
\begin{equation}
\label{eq:LoI}
    \mathscr{L}(A)=
    \begin{cases}
        \mathscr{D}\log_{10}(A)+c_0 & 0<A<A_{\mathrm{pivot}}\\
        \mathscr{L}_0 & A_{\mathrm{pivot}}<A<10^{10}\\
    \end{cases}
\end{equation}
$\mathscr{L}_0$ is the likelihood of intactness of a given system after a predefined age $A_\mathrm{pivot}$ after which disruption may not occur. $c_0$ is the y-value where the sloped piece of the function crosses $A = 1$ yr, but it represents no physical quantity in our simulations because we assume that systems are born at $A = 10^{6}$ yr. We can derive $c_0$ from the other three independent parameters which define the function, $\mathscr{D}$, $\mathscr{L}_0$, and $A_\mathrm{pivot}$:
\begin{equation}
\label{eq:c_0}
    c_0=\mathscr{L}_0-\mathscr{D}\log_{10}(A_{\mathrm{pivot}})
\end{equation}
 
 We give the different values of $\mathscr{D}$, $c_0$, $\mathscr{L}_0$, and $A_\mathrm{pivot}$ for our 9 functions in Table \ref{table:LoI}. The 9 functions differ from one another by 2 quantities, decay rate $\mathscr{D}$ and total intact fraction $\bar{\mathscr{L}}$ where
\begin{equation}
\label{eq:D}
    \mathscr{D}=\frac{d\mathscr{L}}{d\log_{10}(A)}
\end{equation}
and
\begin{equation}
\label{eq:L_bar}
    \bar{\mathscr{L}}=\frac{1}{10^{10}-10^{6}} \int_{10^6}^{10^{10}}\mathscr{L}(A)dA
\end{equation}
$\mathscr{D}$ and  $\bar{\mathscr{L}}$ are also provided in Table \ref{table:LoI} along with their approximate values $\approx\mathscr{D}$ and  $\approx\bar{\mathscr{L}}$.
Hereafter, we reference only the approximate values, which serve to classify each function into one of 9 combinations.

\begin{table}[ht!]
\centering
 \begin{tabular}{||c c c c c c c||}
 \hline
 $\approx\bar{\mathscr{L}}$ & $\approx\mathscr{D}$ & $\bar{\mathscr{L}}$ & $\mathscr{D}$ & $c_0$ & $\log_{10}(A_{\mathrm{pivot}})$ & $\mathscr{L}_0$ \\
 \hline\hline
 0.1 & slow & 0.083 & -0.14 & 1.4 & 9.7 & 0.052 \\
 \hline
 0.1 & medium & 0.097 & -0.14 & 1.4 & 9.1 & 0.09 \\
 \hline
 0.1 & fast & 0.096 & -0.22 & 1.8 & 7.8 & 0.095 \\
 \hline
 0.2 & slow & 0.211 & -0.11 & 1.2 & 9.0 & 0.21 \\
 \hline
 0.2 & medium & 0.226 & -0.14 & 1.4 & 8.0 & 0.23 \\
 \hline
 0.2 & fast & 0.209 & -0.22 & 2.3 & 9.8 & 0.15 \\
 \hline
 0.4 & slow & 0.415 & -0.12 & 1.5 & 9.8 & 0.38 \\
 \hline
 0.4 & medium & 0.418 & -0.18 & 2.1 & 9.0 & 0.41 \\
 \hline
 0.4 & fast & 0.451 & -0.23 & 2.2 & 7.8 & 0.45 \\
 \hline
\end{tabular}
\caption{Piecewise LoI functions defined for each total intact fraction and decay rate combination.}
\label{table:LoI}
\end{table}

Each disrupted system is also assigned an ``age when disrupted'' $A_{d}$ based on the same LoI functions and the same random number $r$ assigned previously:
\begin{equation}
    A_d = \mathscr{L}^{-1}(r).
\end{equation}
Systems that form dynamically warmer, in what we have called the ``disrupted" state, are assigned a default $A_{d}$ of 10$^{6}$ years.  The ``evolutionary timescale'', the ``quiescent" age of the system in which no subsequent sculpting has occurred, is designated $\tau_\mathrm{sys}$:

\begin{equation}
    \tau_\mathrm{sys}=
    \begin{cases}
        A & \mathrm{Intact\ Systems}\\
        A - A_{d} & \mathrm{Disrupted\ Systems}\\
    \end{cases}
\end{equation}

Thus, we have a simplified dynamical narrative for each planetary system. It was formed $A$ years ago, in either a dynamically cold (``intact") or warmer (''disrupted") state. Of the systems born ``intact", some become disrupted over their lifetimes. The disruption event occurred at time $A_{d}$ between birth and its current age. We emphasize here that this is a toy model for dynamical sculpting. We understand that, in reality, dynamical sculpting could operate as a series of events over longer timescales. The time since disruption, in which the system is presumed to have ceased dynamically evolving, is $\tau_\mathrm{sys}$. Due to our assumption of a single disruption event, the values that we find for $\tau_\mathrm{sys}$ are maximum possible values. We refer to this quantity, for the sake of our experiment, as the ``evolutionary timescale''; that is, the period of contiguous time in which hypothetical organisms on the surface of planets in the system have not endured a disruption event. As an example, in a universe with ``medium" speed dynamical sculpting: a 10-Gyr-old star that was born dynamically cold might hypothetically disrupt $A_{d}=1$ Gyr into its life. In its disrupted state, it has proceeded without further orbital changes for $\tau_\mathrm{sys}=9$ Gyr. 

To begin assembling the planets in each system, we draw on occurrence rate calculations from \cite{Dressing15}, \cite{Muirhead15}, and \cite{Ballard16}. Based on its pre-assigned dynamical state (determined from the age of the star), a system is assigned a number of planets $N_{p}$ and scatter in mutual inclinations $\sigma_{i}$ chosen from one of two posterior distributions published in the supplemental data of \cite{Ballard16}. One of these distributions in $\{N_{p},\sigma_{i}\}$ space corresponds to intact systems, and one to disrupted systems. Disrupted systems have between 1 and 3 planets, with mutual inclinations of several degrees, while intact systems have between 4 and 8 planets, with mutual inclinations less than 2$^{\circ}$. 

Before assigning individual planet properties, we sample a mean orbital plane inclination $\bar{\theta}$ from a uniform distribution:
\begin{equation}
    \bar{\theta} \sim \mathscr{U}(-90,90^{\circ})
\end{equation}
The angle $\bar{\theta}$ is measured relative to an observer's line of sight, so $\bar{\theta}=0$ is edge-on from the observer's perspective.

\subsection{Planet Properties}
\label{sec:Planet-Properties}

\begin{figure}[b!]
\centering
\includegraphics[width=3.3in]{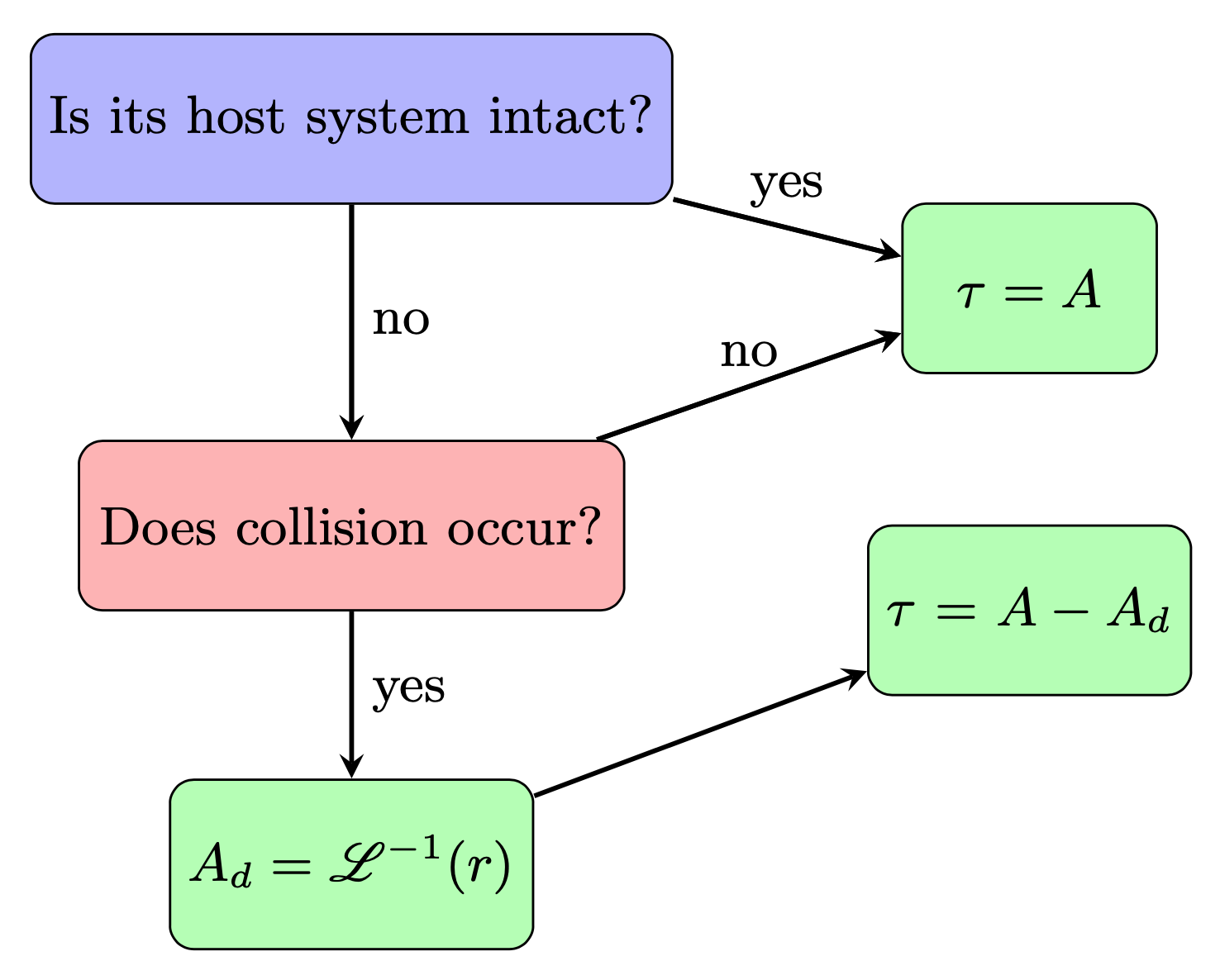}
\caption{Flowchart showing how evolutionary timescale $\tau$ of a planet is chosen.}
\label{flowchart}
\end{figure}

We store planet properties as lists of different lengths (where the length is the number of planets) for each system. In intact systems, the evolutionary timescale of each planet $\tau$ is equal to the system age. To compute the evolutionary timescale $\tau$ for each planet in disrupted systems, we first need to assign them \textit{disruption modes}. A ``disruption event" for our purposes is one that results in a dynamically warmer configuration; such an event could be as catastrophic as a planet-planet collision (such as the late-heavy bombardment event leading to the unusual composition and orbit of Mercury). Or, it might manifest in a way that poses a less immediate threat to hypothetical organisms, such as a modest increase in eccentricity occurring over millions of years. We investigate both scenarios, where \textit{collision} is a disruption mode in which no organisms persist and the evolutionary clock resets to zero ($\tau=A-A_d$). Alternatively, \textit{migration} is a disruption mode in which the evolutionary clock need not reset to zero, and we treat the evolutionary timescale $\tau$ for a planet as that of a planet in an intact system, that is $\tau=A$. We include a parameter called the \textit{collision fraction} $f_{c}$ in our simulations, which simply determines the probability that any planet in a disrupted system undergoes a catastrophic event in which all putative life become extinct. Throughout this paper, we assume $f_{c}=0.5$ unless otherwise noted. We make this choice for simplicity given that we have no prior knowledge of the implications of disruptive events on evolutionary clocks. We therefore assume that disruptive events cause evolutionary clocks to reset half of the time in our simulations. The implication of this choice for our results is that the mean $\tau$ of planets in disrupted systems is an intermediate value between that of full collision (always reset evolutionary clock) and full migration (never reset). We explore the effects of varying $f_{c}$ in Section \ref{sec:Collison-Fraction}. This process by which $\tau$ is assigned is shown in Figure \ref{flowchart}. This method by which we assign the parameters that determine $\tau$ is such that we can numerically predict the distributions of $\tau$ in disrupted and intact systems. The formalism for these numerical predictions is provided in the Appendix in Section \ref{sec:Numerical-Predictions}.

We choose orbital periods $P$ of planets log-uniformly per \cite{Foreman_Mackey_2014}:
\begin{equation}
    \ln\left(\frac{P}{\mathrm{day}}\right) \sim \mathscr{U}\left(\ln(0.75), \ln(300)\right)
\end{equation}
Semi-major axes $a$ are then calculated using Kepler's third law from the periods and $M_{\star}$.

Radii and masses of planets are randomly drawn from those of a mock sample of planets orbiting \textit{TESS} targets from \cite{Ballard19}.

Inclinations $\theta$ of planets are chosen from Gaussian distributions,
\begin{equation}
    \theta \sim \mathscr{N}(\bar{\theta}, \sigma_{i})
\end{equation}
where the standard deviation $\sigma_{i}$ is chosen from \cite{Ballard16} as previously described in Section \ref{sec:System-Properties}.

Eccentricities $e$ of planets are chosen from two different probability distributions, as well. Rather than assuming a standard relationship between  $\sigma_{e}$ and $\sigma_{i}$ and assigning $\sigma_{e}$ directly from $\sigma_{i}$, we elected to draw eccentricities from their observed distributions for ``single" and ``multiple" transiting planet systems from \cite{VanEylen_2019}: Disrupted systems have $e$ drawn from a Rayleigh distribution, and intact systems have $e$ drawn from a half-Gaussian distribution.
\begin{equation}
    e=
    \begin{cases}
        |\mathscr{N}(0,\sigma=0.049)| & \mathrm{Intact\ Systems}\\
        \mathscr{R}(\sigma=0.26) & \mathrm{Disrupted\ Systems}\\
    \end{cases}
\end{equation}

\begin{figure*}
    \centering
    \includegraphics[trim={0.5cm 0 0.5cm 0}, width=7.0in]{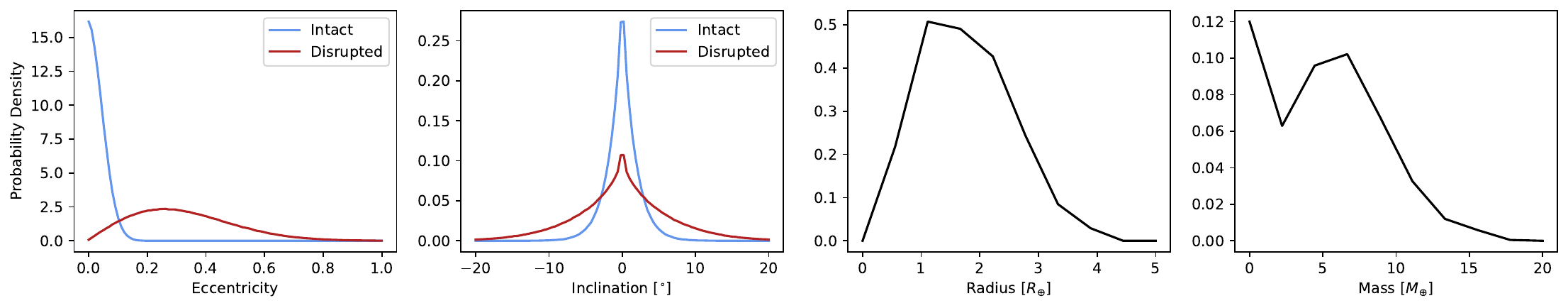}
    \caption{Histograms of the distributions from which planet properties are chosen. Eccentricity and inclination of a planet depend on the dynamical state of its host system, while radius and mass do not. Note that inclinations $i$ shown here are relative to the mean orbital plane inclination of each system.}
    \label{fig:Planet-Properties-Histograms}
\end{figure*}

We show the distributions of eccentricity, inclination, radius, and mass in Figure \ref{fig:Planet-Properties-Histograms}. Note that the inclination displayed here is the relative inclination $i$ which is found by subtracting the mean orbital plane inclination $\bar{\theta}$ of a planet's host system from the inclination $\theta$ of the planet. 

\begin{figure}
    \centering
    \includegraphics[width=3.4in]{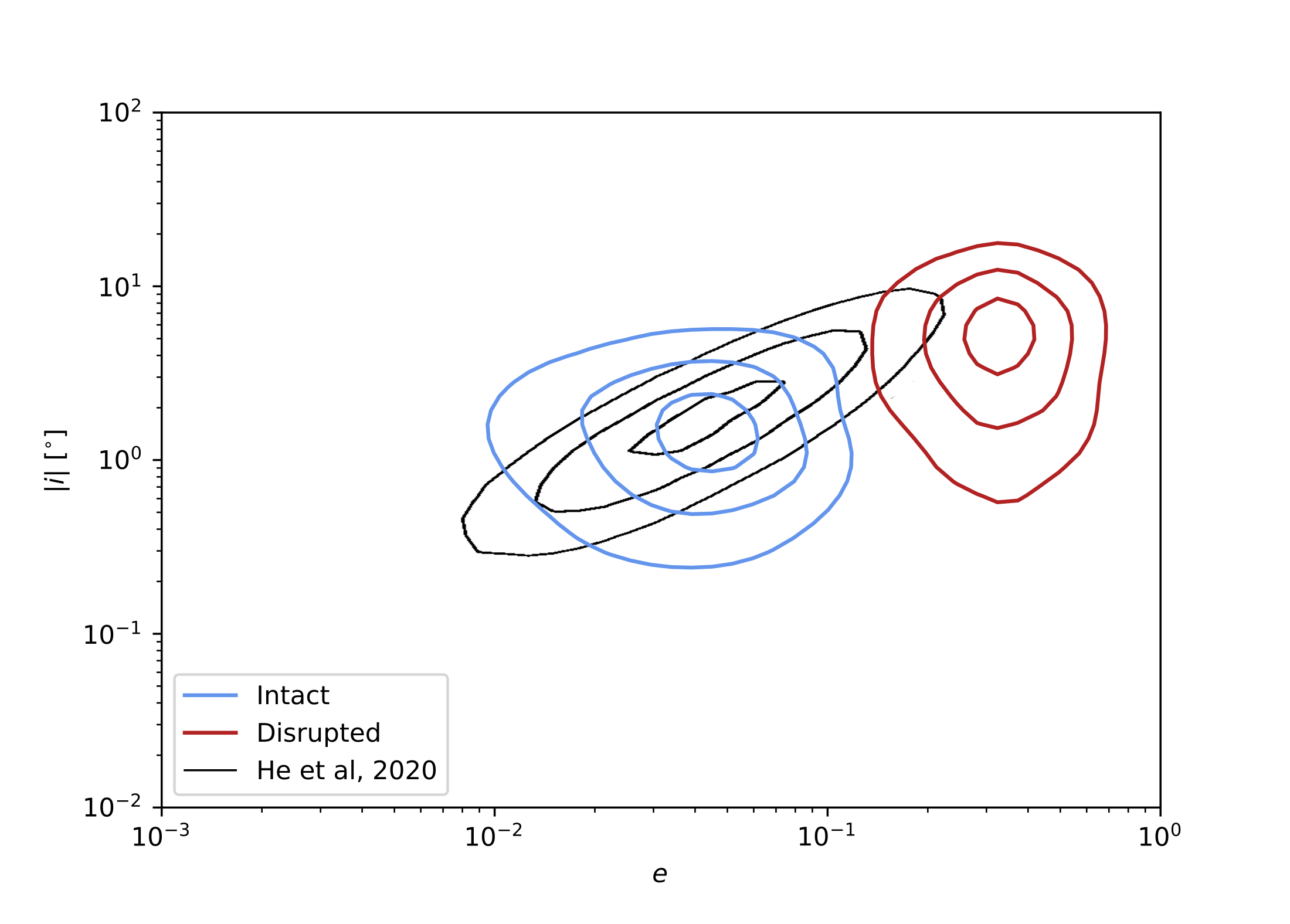}
    \caption{Contour plots of inclination $i$ versus eccentricity $e$ distributions that we draw from compared to that of ``intrinsic multi-planet systems” from the maximum AMD model of \cite{He20}}
    \label{fig:Inclinations-Vs-Eccentricities}
\end{figure}

Contours of $|i|$ versus $e$ for both ``intact" and ``disrupted" contributions to our planetary sample are plotted in Figure \ref{fig:Inclinations-Vs-Eccentricities}, against the ``maximum AMD'' model of \cite{He20} for comparison. We note here the broad consistency between the $\{e,i\}$ distribution from the theoretical AMD stability limit, and the observationally-derived $\{e,i\}$ distributions that we have employed. The disrupted and intact systems predictably cluster at the high and low AMD ends of $|i|$ versus $e$ parameter space. 

We choose longitudes of periastron $\omega$ of planets from a uniform distribution:
\begin{equation}
    \omega \sim \mathscr{U}(0,360^{\circ})
\end{equation}

Per \cite{Kopparapu_2013}, we define a planet to be ``in the habitable zone'' if it meets the following condition:
\begin{equation}
    0.2<a<0.38\ \rm{AU}
\end{equation}
This condition (specifically for a $0.5 M_{\odot}$ star) constrains the habitable zone by loss of water at the inner edge and the maximum greenhouse provided by a $\mathrm{CO}_2$ atmosphere at the outer edge.

\subsection{Simulated Observations}
\label{sec:Simulated-Observations}

Given previously assigned planet properties, we calculate the impact parameters $b_{\rm{transit}}$ of planets using the following equation from \cite{winn2014transits}:
\begin{equation}
    b_{\rm{transit}} = a\sin(\theta)\left(\frac{1 - e^2}{1 + e \sin(\omega)}\right)
\label{eq2}
\end{equation}
If $|b_{\rm{transit}}|<R_{\star}$, the planet is assigned to ``transit" from the perspective of a hypothetical observer.

\begin{figure*}
    \centering
    \includegraphics[trim={4cm 0 4cm 0}, width=7.0in]{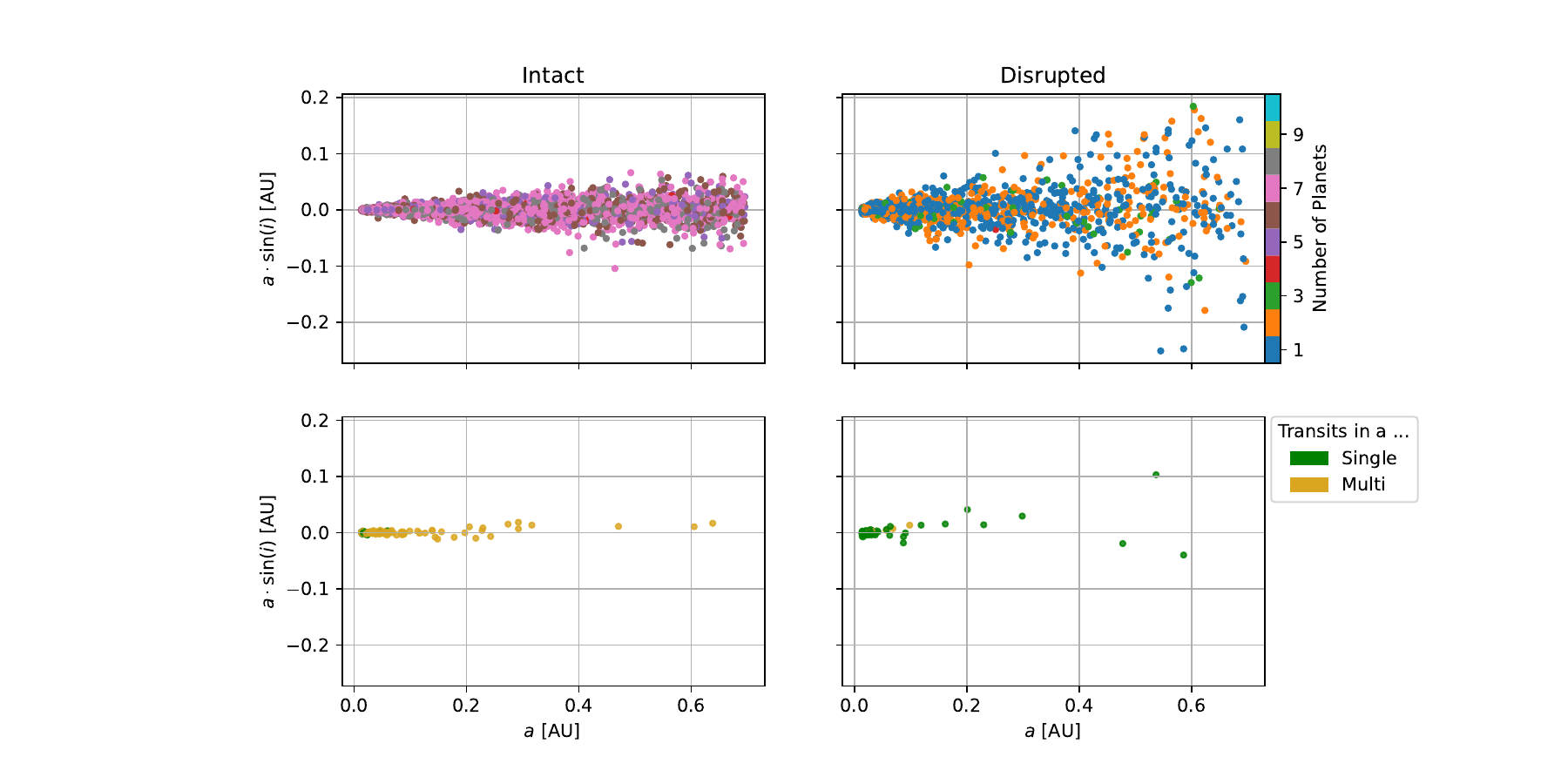}
    \caption{Scatter plots of $a\cdot\sin{i}$ versus $a$ for planets from a simulation ($\bar{\mathscr{L}}=0.1, \mathscr{D}=$ slow) with 2500 systems. The top panels show all planets in the simulation, colored by the number of planets in their host system. The bottom panels show only planets which transit, and are colored according to whether they do (multi) or do not (single) have transiting companions.}
    \label{fig:Typical-Systems}
\end{figure*}

Figure \ref{fig:Typical-Systems} shows all of the planets from a simulation with 2500 systems, then isolates those that transit as seen by the hypothetical perfect observer. By displaying $a\cdot\sin{i}$ versus $a$ for each planet, the difference in the spread of inclinations between intact and disrupted systems becomes visible. For stars hosting intact systems, a substantial fraction of the time that one planet transits, there are additional transiting planets as well \citep{Ballard16}. This is due to both the higher number of planets and the lower mutual inclinations in intact systems. Conversely, planets in disrupted systems usually transit without companions.

Finally, we determine which of the synthetic transiting planets in our samples are ``detected" to transit. We elect to model both the detection completeness of the \textit{Kepler} and \textit{TESS} missions. We make the simplifying assumption that \textit{Kepler} detects 100\% of the planets that transit M dwarfs. The completeness function is more complex in reality, per \cite{Dressing15}, but a typical 2$R_{\oplus}$ planet that transits will be detected with high probability for orbital periods as long as 200 days. We cannot apply a similar simplifying assumption to our synthetic \textit{TESS} observations, for which the shorter typical observing window complicates the completeness. While \textit{TESS}' exact completeness to M dwarf planets has not yet been empirically measured, we employ the modeled completeness function from \cite{Ballard18} (that is, we draw a representative sample of completeness functions from that work, to marginalize over the uncertainty about the completeness). We use the completeness function to assign a probability that a planet is detected by \textit{TESS}, based on the planet radius and period.

\begin{figure*}
    \centering
    \includegraphics[width=7.0in]{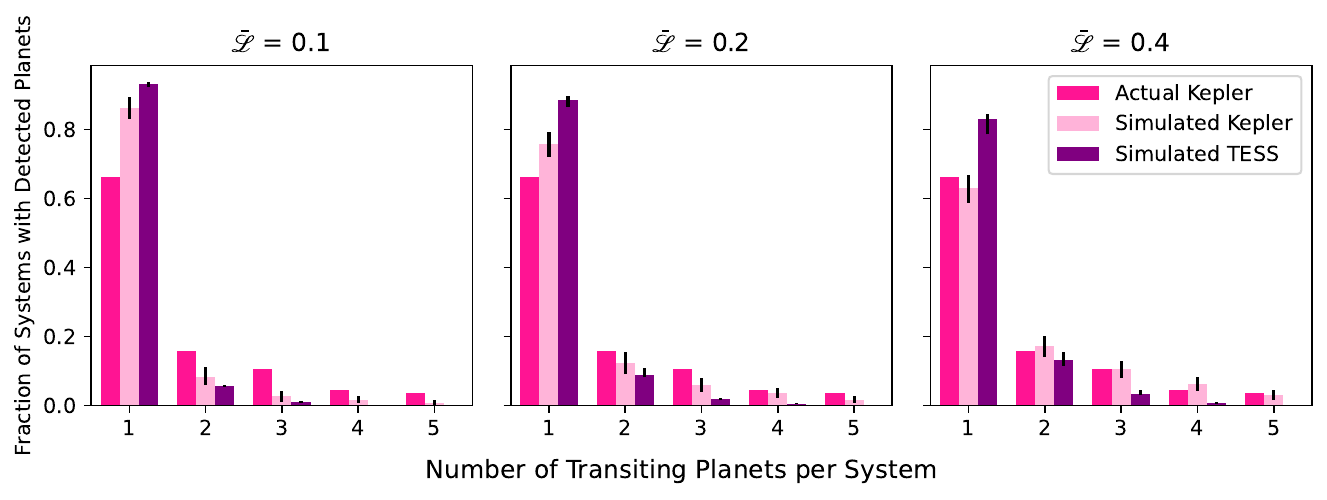}
    \caption{Bar plots of the fraction of planet hosts in a simulation which have a given number of transiting planets. Actual data from \textit{Kepler} and simulation results from both \textit{Kepler} and \textit{TESS} are shown. The $+/-$ error shown here is the standard deviation of fraction of systems over the collection of simulations.}
    \label{fig:Number-of-Transits}
\end{figure*}

We now compare our ``observed'' simulation yields to actual mission yields for the sake of validation. To select our real comparison sample, we search the \cite{thompson_2018} catalog for stellar hosts with effective temperatures $<$ 4000 K and disposition scores below 0.5. From this subset, we exclude planets with an Exoplanet Archive Disposition of ``False Positive", according to the tests performed by \citet{batalha_2013}. This sample contains 115 host stars: 76 systems with 1 transiting planet, 18 with 2, 12 with 3, 5 with 4, and 4 with 5. Figure \ref{fig:Number-of-Transits} compares this data to our simulation data by displaying the fraction of systems with detected planets that possess each number of transiting planets between 1 and 5. We see that the model closest to the actual yield corresponds to a total intact fraction (or ``compact multi" rate) of $\bar{\mathscr{L}}=0.4$. This is slightly higher than the compact multi rate of 0.2 inferred among among early M dwarfs determined by both \cite{Muirhead15} and \cite{Ballard18}, a fact we attribute to the way we assign orbital periods to our sample. While \cite{Ballard16} drew planets in uniform log space between 0.5--200 days, we have employed a uniform log space 0.75-300 days, resulting in systems with wider spacings between adjacent planets. However, we see that models with $\bar{\mathscr{L}}=0.1$ and $\bar{\mathscr{L}}=0.2$ are only in modest tension with the observations; they both lie within the 2$\sigma$ confidence interval for $\bar{\mathscr{L}}$. The \textit{TESS} simulated yield shows less multiply-transiting systems as compared to single-transiting systems. This is consistent with the mission's lower completeness to transits with longer orbital periods, given the mission's typical 27-day baseline per star. 

\subsection{Size of Simulated Sample}

The parameters that characterize each simulation are as follows: 
\begin{itemize}
    \item Total intact fraction ($\bar{\mathscr{L}}$): Probability that any system is intact
    \item Decay rate ($\mathscr{D}$): Rate at which systems become disrupted
    \item Collision fraction ($f_c$): Probability that any planet in a disrupted system has its evolutionary clock reset
    \item Number of simulated systems ($N$)
\end{itemize}

We perform simulations for each of the 9 $\{{\mathscr{L}}$,$\mathscr{D}\}$ combinations given in Table \ref{table:LoI}. We keep $f_c$ fixed at 0.5 in each of these simulations (though we consider the implications of this assumption in Section \ref{sec:Collison-Fraction}). 

For each $\bar{\mathscr{L}}$, $\mathscr{D}$ combination, we generate and store three sets of simulations: one large enough to sample the ``inherent" properties of planetary systems in the galaxy, one that simulates the subsample of these systems observed by \textit{Kepler}, and one that simulates the subsample of these systems observed by \textit{TESS}. Accordingly, our inherent simulations use $N=10^6$ stars, our  \textit{Kepler} simulations use $N=2500$ \citep{Dressing15}, and our \textit{TESS} simulations use $N=70000$ \citep{Sullivan15,Muirhead17}. To acquire approximately the same total number of simulations ($10^6$) in each set, we run 1 inherent simulation, 400 \textit{Kepler} simulations, and 15 \textit{TESS} simulations. It is worth noting that an alternative way to obtain the same data is to run a single set of $10^6$ simulations and then break it into the \textit{Kepler}- and \textit{TESS}-sized samples, applying the completeness corrections to each, respectively.

\section{Results}
\label{sec:Results}

In this Section, we consider the results of our simulated samples. Given the number of assumptions, we elect to present results for only the limiting cases of $\bar{\mathscr{L}}$ and $\mathscr{D}$. This comprises four suites of simulations, in which the total intact fraction is either 0.1 or 0.4 and the dynamical sculpting is either ``slow" or ``fast".

In Section \ref{sec:Transit-Demographics}, we consider the way that intact and disrupted systems contribute to the predicted observables for \textit{Kepler} and \textit{TESS}. In Section \ref{sec:Evolutionary-Timescales} we investigate how our assumptions for dynamical sculpting map to the resulting evolutionary timescales for the population of M dwarf planetary systems. We go on in Section \ref{sec:Angular-Momentum-Deficit} to consider how the predicted movement in $\{e,i\}$ space for our dynamical sculpting laws manifests as increasing average angular momentum deficit over long timescales. In Section \ref{sec:Collison-Fraction} we describe how varying the catastrophic collision fraction among disrupted systems affects our findings. And finally, in Section \ref{sec:Habitability} we examine the predicted evolutionary timescales among the set of ``habitable" planets. 

\subsection{Transit Multiplicity Demographics}
\label{sec:Transit-Demographics}

We assess our ability to employ transit multiplicity (``multi" versus ``single") as a proxy for dynamical state (intact versus disrupted). To first order, we know these quantities ought to be correlated. In dynamically cooler, intact systems, planets reside in more closely spaced configurations with low mutual inclinations, and thus ought to result in more ``multis", where two or more transiting planets are observed. Planets in dynamically hotter, disrupted systems will be more mutually inclined and less likely to host more than one transiting planet, thus producing ``singles'', where only one transiting planet is observed.

\begin{figure*}
    \centering
    \includegraphics[width=7.0in]{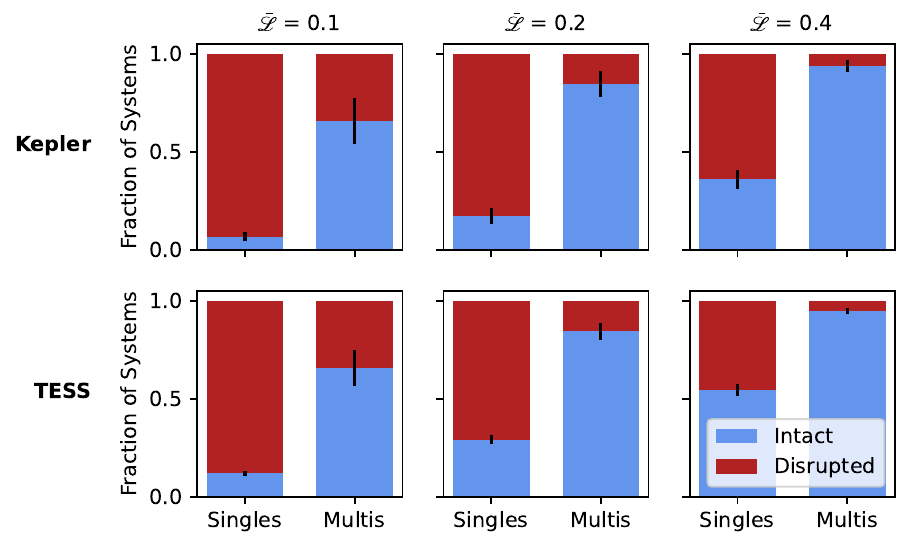}
    \caption{The fraction of singly- and multiply-transiting systems in each simulation which are either intact or disrupted. This is shown for the three intrinsic intact rates  $\bar{\mathscr{L}}$ for both the \textit{Kepler} and \textit{TESS} simulations. The $+/-$ error shown here is the standard deviation of intact fraction over the collection of simulations.}
    \label{fig:Fraction-of-Observed-Systems-By-Disruption-State}
\end{figure*}

We analyze this correlation in our own \textit{Kepler} and \textit{TESS} simulated observations in Figure \ref{fig:Fraction-of-Observed-Systems-By-Disruption-State}. Here, we show the fraction of singles and multis which are intrinsically intact and disrupted systems. We see that the correlations between transit multiplicity and dynamical state depend both on $\bar{\mathscr{L}}$ and the mission completeness. Consider, for a moment, only the intact systems, shown in blue. The population of singles will contain more intact systems (1) when the mission completeness is lower (as in \textit{TESS}, when additional planets may transit but are not detected) and (2) when the intrinsic rate of intact systems is higher. In all cases, the fraction of singles which are intact approximates the assigned $\bar{\mathscr{L}}$. The fraction of multis which are intact is greater than 50\% in all cases. These results along with those from Figure \ref{fig:Typical-Systems} show that intact systems map closely to multis, and disrupted systems are likely to be singles. 

In Figure \ref{fig:Fraction-of-Observed-Systems-By-Disruption-State}, intact systems comprise a fraction of detections that is larger than the intrinsic intact fraction. This phenomenon stems from the geometric transit probability ($R_{\star}/a$) which mandates that intact systems, which host more planets, including in close-in orbits, are likeliest to produce a transit. Per \cite{Ballard18}, if the intrinsic intact fraction is $\bar{\mathscr{L}}=0.2$, \textit{Kepler} will disproportionately yield 50\% intact systems in its surveys, and more extreme, \textit{TESS} will yield 70\%. These estimates are roughly reproduced.

\begin{figure*}
    \centering
    \includegraphics[width=7.0in]{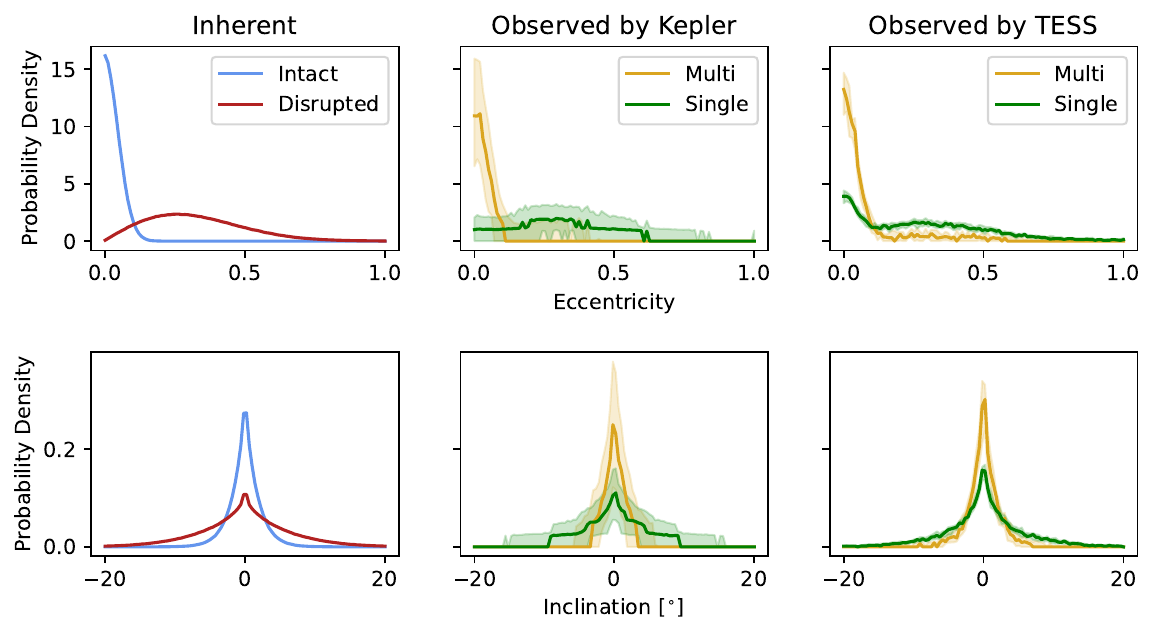}
    \caption{Histograms of eccentricity $e$ and inclination $i$ divided into subgroups: intact and disrupted from inherent samples and multi and single from observed samples (using $\bar{\mathscr{L}}=0.1$, $\mathscr{D}=$ slow). The median of probability density along with an error contour containing 68\% of the data from \textit{Kepler} and \textit{TESS} is plotted.}
    \label{fig:Eccentricity-and-Inclination-Histograms}
\end{figure*}

Finally, it is useful to examine the distributions of eccentricity and inclination among the inherent and the synthetic detected samples. Using $\bar{\mathscr{L}}=0.1$, $\mathscr{D}=$ slow, Figure \ref{fig:Eccentricity-and-Inclination-Histograms} displays histograms of $e$ and $i$ for inherent samples, subdividing by dynamical state, and for \textit{Kepler} and \textit{TESS} samples, subdividing by transit multiplicity. As in Figure \ref{fig:Fraction-of-Observed-Systems-By-Disruption-State}, it is clear that these are useful but imperfect proxies, with intact mapping to multis in a cleaner fashion than disrupted maps to singles. 

\subsection{Evolutionary Timescales}
\label{sec:Evolutionary-Timescales}

\begin{figure*}
    \centering
    \includegraphics[width=7.0in]{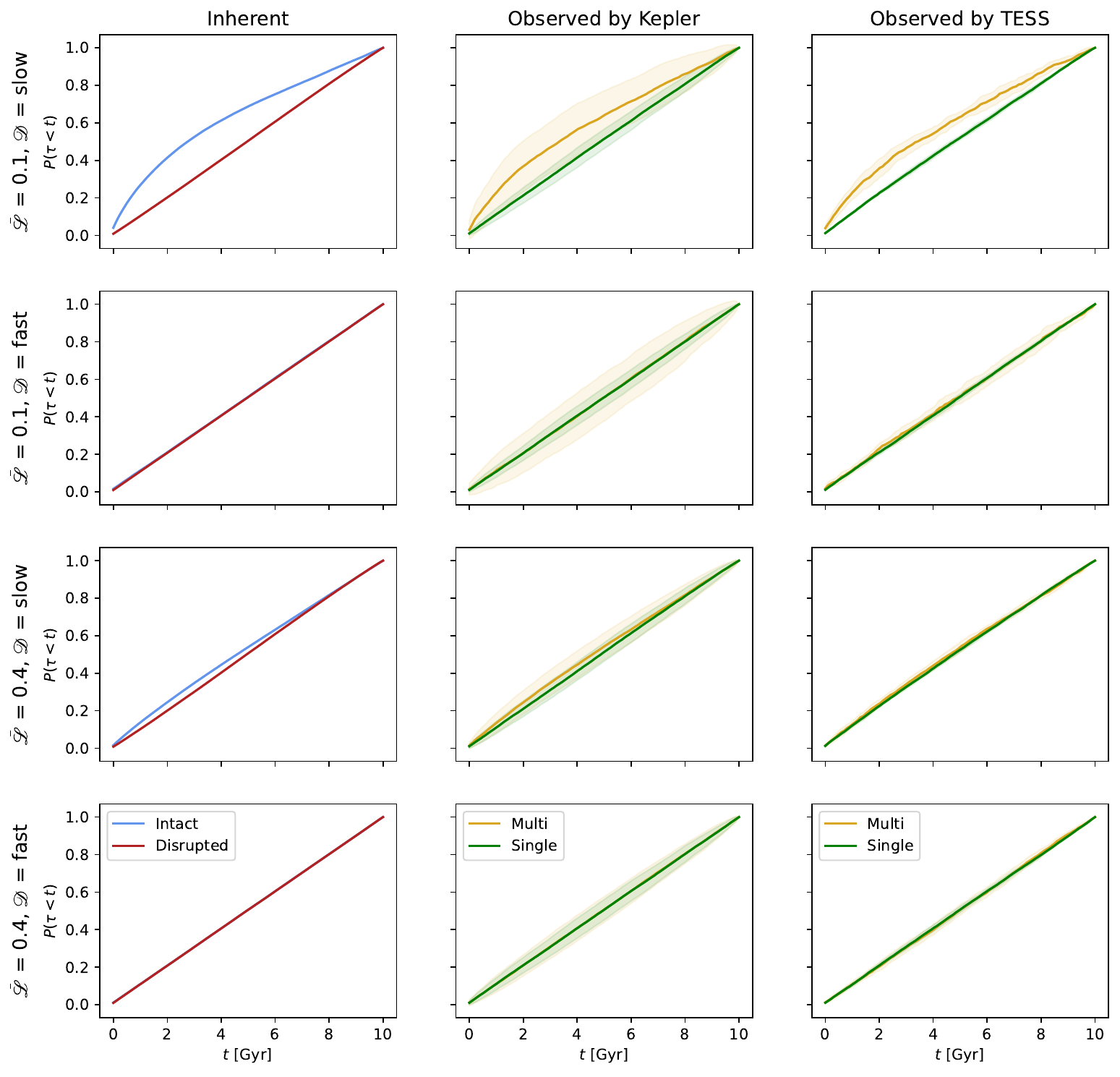}
    \caption{Cumulative histograms of evolutionary timescale of planets $\tau$ in the inherent and observed samples for the limiting cases of $\bar{\mathscr{L}}$ and $\mathscr{D}$. The $+/-$ error shown here is the standard deviation over the collection of simulations.}
    \label{fig:Tau-Histogram}
\end{figure*}

We next explore the distributions of planet evolutionary timescale $\tau$ in our simulations. We have posited hypothetical dynamical sculpting over long timescales, with systems moving from intact to disrupted with some probability as they age. We hypothesize that the disruption process, particularly if it induces planet-planet collision, could ``reset" the evolutionary clock. If the $\tau$ distributions differ between subgroups, this could imply that life, if it exists, has had more time to evolve in one of the subgroups. Figure \ref{fig:Tau-Histogram} shows the cumulative distributions of $\tau$ for the intrinsic and observed samples for our limiting cases of $\bar{\mathscr{L}}$ and $\mathscr{D}$. Note that in this analysis, we include all planets, not solely those which are assigned habitable according to our definition. We explore in Section \ref{sec:Habitability} how applying our specific habitability criterion may affect the distributions.

We can now compare the relative contributions of two competing effects. We might expect (1) that $\tau$ should be \textit{higher} on average for intact systems. This is because disrupted systems have undergone a disruption and ``reset" event, shortening their $\tau$. Dependent on the collision fraction, $\tau=A-A_{d}$ rather than $A$ for disrupted systems. Intact systems, in contrast, never experienced such a reset, and their $\tau=A$ always. Alternatively, (2) $\tau$ might be \textit{lower} for intact systems. This effect would be attributable to the youthfulness of intact systems generally. While they have undergone no disruption event, their very intactness means that the star is younger and less time has elapsed on the surface.

Firstly, we find we can distinguish between these cases only for ``slow" sculpting scenarios. When sculpting occurs early (i.e. $\mathscr{D}=$ fast), the distributions of the two subgroups are nearly identical. This is because hypothetical disruption occurred so soon after formation, the resulting quiescent period is effectively the age of the star. We only see a difference in $\tau$ for intact and disrupted systems when sculpting occurs later into the stellar life. We find that effect (2) is much stronger than (1), to the extent that intact/multi systems exhibit observably \textit{lower} $\tau$ lower values. Stated differently, while it is true that $\tau$ has been shortened by $A_{d}$ for disrupted systems, this effect is overwhelmed by the fact that these systems are necessarily already older. As an example, for slow sculpting resulting in an intact fraction $\bar{\mathscr{L}}=0.1$, only $\sim$50\% of stars hosting intact systems have $\tau \ge$3 Gyr. This is in contrast with disrupted systems, for which $\sim$75\% have $\tau \ge$3 Gyr.

This effect is greater for $\bar{\mathscr{L}}=0.1$ than $\bar{\mathscr{L}}=0.4$. This is because, given the lower occurrence of disruption among systems in a $\bar{\mathscr{L}}=0.4$ scenario, systems are likelier to remain intact to old age. This decreases the extent to which disrupted systems are \textit{a priori} so much older than intact systems as to offset the effect of a disruptive ``reset."  

However, this interpretation is only valid when considering the entire sample of stars. In this case, it is clear from Figure \ref{fig:Tau-Histogram} that $\tau$ is greater in disrupted/single systems given $\mathscr{D}=$ slow. However, this changes if the observer has age information about the host star. If we are able to identify samples of stars older than a given age $A$, the relative $\tau$ properties change. 

\begin{figure*}
    \centering
    \includegraphics[width=7.0in]{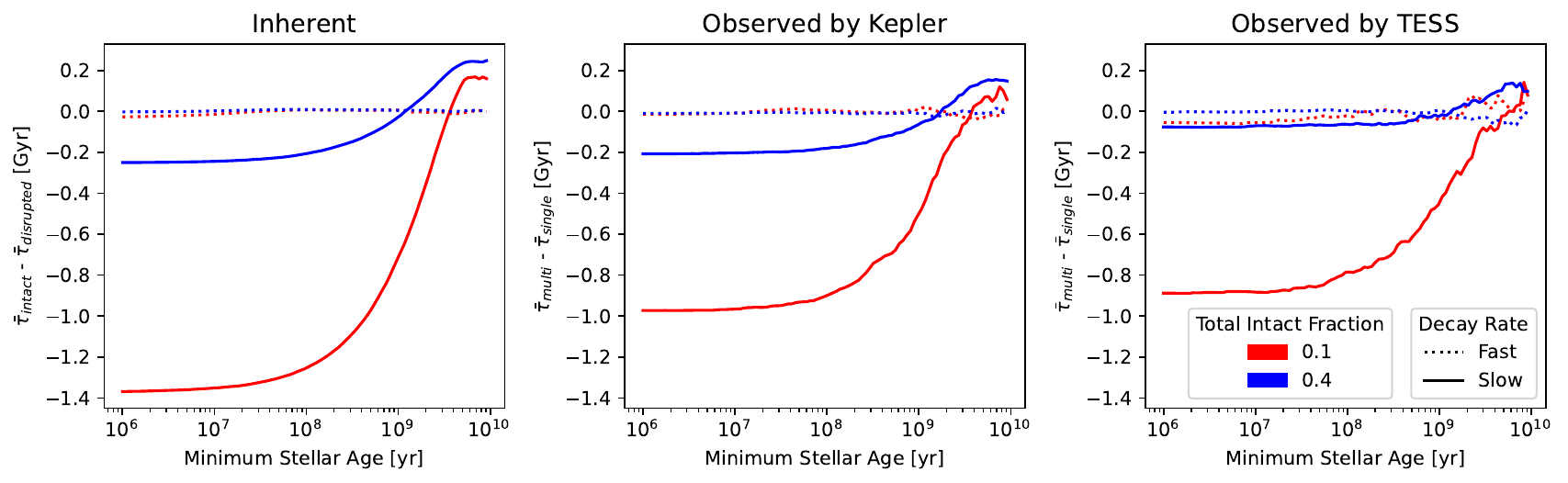}
    \caption{The difference between the mean intact/multi $\tau$ and the mean disrupted/single $\tau$ ($\bar{\tau}_{\textrm{intact/multi}} - \bar{\tau}_{\textrm{disrupted/single}}$) at different minimum stellar ages $A_\textrm{min}$ considered for the limiting combinations of $\bar{\mathscr{L}}$ and $\mathscr{D}$.}
    \label{fig:Delta-Tau-Vs-Min-Age}
\end{figure*}

We consider a quantity $\Delta\tau$ equal to the difference between the mean intact/multi $\tau$ and the mean disrupted/single $\tau$ ($\bar{\tau}_{\textrm{intact/multi}} - \bar{\tau}_{\textrm{disrupted/single}}$). This quantity is negative for the sample as a whole, because planets in disrupted systems have longer $\tau$ as discussed above. However, there exists an age, where $\Delta\tau$ becomes positive among a sample of systems all older than that age. That is, if a planet in an intact system has survived to that age, it likely has a longer $\tau$ than a planet in a disrupted system of that age. Figure \ref{fig:Delta-Tau-Vs-Min-Age} shows how this $\Delta\tau=\bar{\tau}_{\textrm{intact/multi}} - \bar{\tau}_{\textrm{disrupted/single}}$ function of minimum stellar age $A_{\textrm{min}}$ behaves when we vary $\bar{\mathscr{L}}$ and $\mathscr{D}$. At the leftmost boundary of these plots, the entire sample is included. As we shift to higher $A_{\textrm{min}}$, systems with younger ages are excluded and the advantage conferred on the $\tau$ of planets in disrupted systems by their longer age diminishes. At the rightmost boundary of these plots, for $\mathscr{D}=$ slow, mean intact/multi $\tau$ is greater than mean disrupted/single $\tau$. The switch from negative to positive $\Delta\tau=\bar{\tau}_{\textrm{intact/multi}} - \bar{\tau}_{\textrm{disrupted/single}}$ occurs at different $A_{\textrm{min}}$ for different $\bar{\mathscr{L}}$ values, specifically around $A_{\textrm{min}}=3\cdot10^9$ for $\bar{\mathscr{L}}=0.1$ and $A_{\textrm{min}}=10^9$ for $\bar{\mathscr{L}}=0.4$. Therefore, focusing on old M dwarf systems, planets in intact/multi systems would have $\tau$ greater than or equal to than those in disrupted/single systems. 

\subsection{Angular Momentum Deficit}
\label{sec:Angular-Momentum-Deficit}

\begin{figure*}
    \centering
    \includegraphics[width=7.0in]{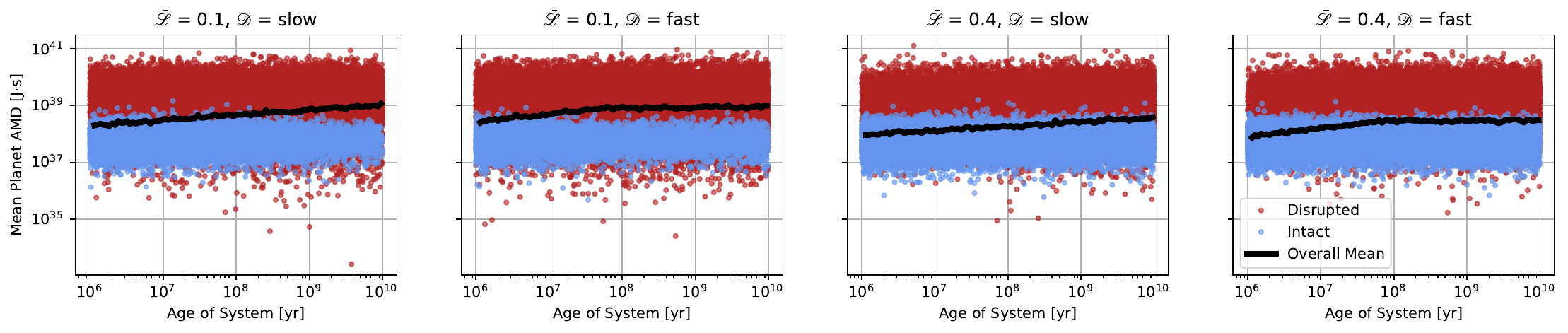}
    \caption{Scatter plots of the mean angular momentum deficit (AMD) of planets in each system from a simulation of $10^6$ systems. Systems are labeled according to their dynamical state, but the mean of \textit{all} systems is plotted.}
    \label{fig:Age-Vs-AMD}
\end{figure*}

Recent papers have evaluated dynamical stability according to angular momentum deficit (AMD). As explained in Section \ref{sec:System-Properties}, we expect that AMD is higher on average for planets in disrupted systems. This is what we see in Figure \ref{fig:Age-Vs-AMD}, which displays the mean planet AMD of $10^6$ systems with log-spaced ages and their dynamical states. In Figure \ref{fig:Age-Vs-AMD}, the mean planet AMD in some disrupted systems reaches down to the same values as those in intact systems because it is possible for planets in those systems to have low $e$ and $|i|$ as shown in Figure \ref{fig:Planet-Properties-Histograms}.
Since more disrupted systems exist with higher ages, we see the mean planet AMD increases as $A$ increases. This follows the different LoI functions shown in Figure \ref{fig:LoI}.

\subsection{Collision Fraction}
\label{sec:Collison-Fraction}

\begin{figure*}
    \centering
    \includegraphics[width=7.0in]{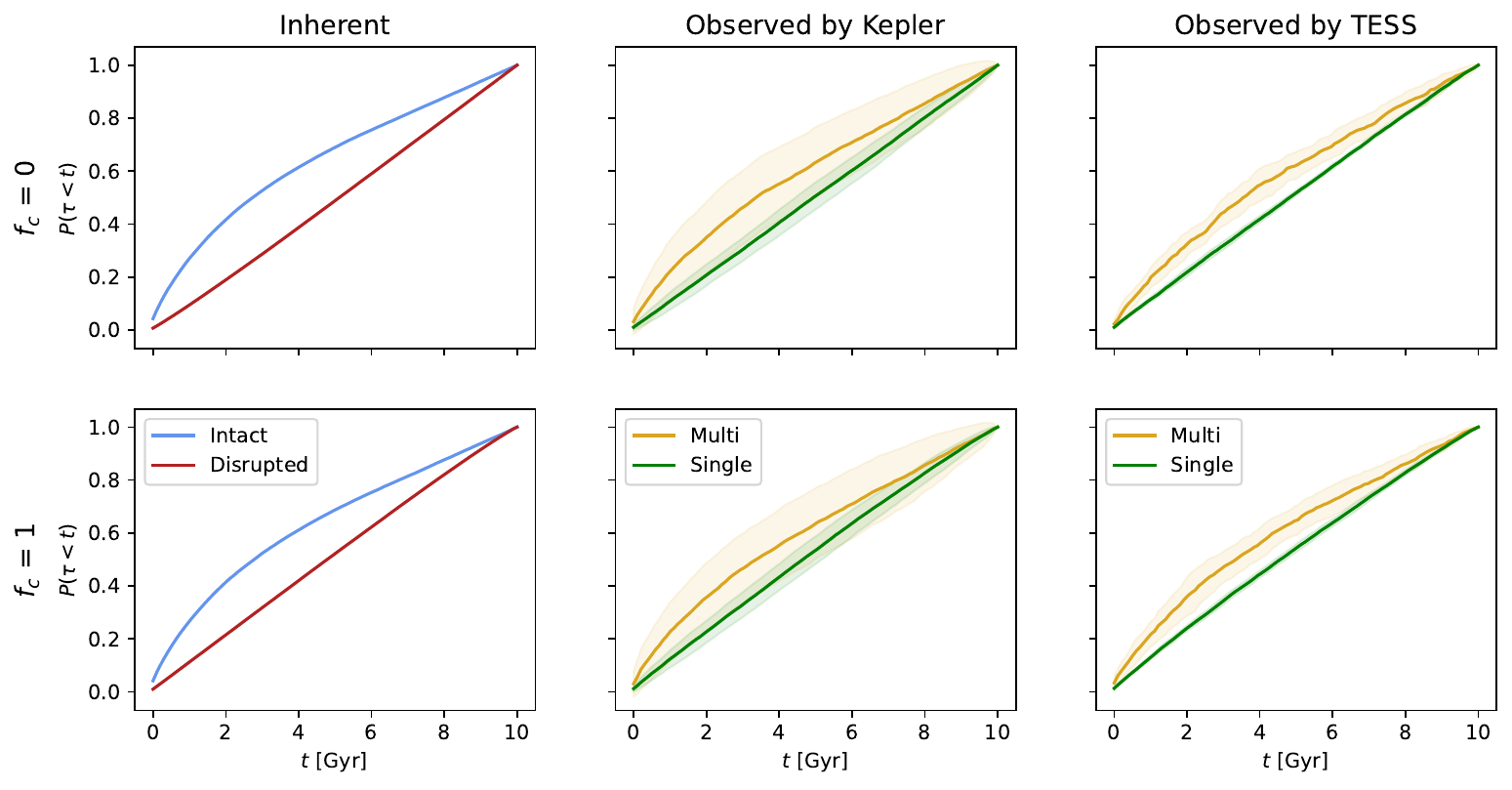}
    \caption{Cumulative histograms of evolutionary timescale $\tau$ in the inherent and observed samples for $f_c=0$ and $f_c=1$ given $\bar{\mathscr{L}}=0.1$ and $\mathscr{D}$=slow. The $+/-$ error shown here is the standard deviation over the collection of simulations.}
    \label{fig:Tau-Histogram-for-Different-CF}
\end{figure*}

We return to the $f_c$ simulation parameter described in Section \ref{sec:Planet-Properties} and Section \ref{sec:Results}. We perform two more sets of simulations, one where $f_c=0$ and another where $f_c=1$. When $f_c=0$, the evolutionary clock of planets in disrupted systems is \textit{never} reset, and the distributions of $\tau$ for planets in intact/multi systems and planets in disrupted/single systems simply reflect their respective age distributions. When $f_c=1$, the evolutionary clock of planets in disrupted systems is \textit{always} reset. This has the effect of reducing the mean $\tau$ for planets in disrupted/single systems. Cumulative histograms (given $\bar{\mathscr{L}}=0.1$ and $\mathscr{D}=$ slow) for these two collision fractions are shown in Figure \ref{fig:Tau-Histogram-for-Different-CF}. These also can be reproduced using Equation \ref{eq:P_tau_in_interval_D} in Section \ref{sec:Numerical-Predictions}.

\subsection{Habitability}
\label{sec:Habitability}

We now consider the $\tau$ distribution among the subsample of ``habitable" planets. Approximately 16\% of planets in any given simulation are habitable; this is true regardless of dynamical state. The habitable zone condition requires that $a$ fall between 0.2 and 0.38 AU; the majority of planets have $a<0.2$ AU. This is due to the way periods are chosen, as described in Section \ref{sec:Planet-Properties}. Because periods are not chosen based on dynamical state, this means that habitability is decoupled from dynamical state, and the habitable planets population possesses the same cumulative histograms shown in Figures \ref{fig:Tau-Histogram} and \ref{fig:Tau-Histogram-for-Different-CF}.

Looking at planets which transit, the portion of those that are habitable decreases to around 3.5\%. This is because planets are more likely to transit at $a$ smaller than 0.2 AU ($P(\mathrm{transit})=\frac{R_{\star}}{a}$). Planets transiting in multi systems have a slightly higher chance ($\sim4\%$) of being habitable than those in single systems ($\sim3\%$). This is due to the differing number of planets and spread in inclinations in those subgroups. If one planet transits at small $a$ in a multi system, then, from coplanarity, one or more transit at higher $a$ which may land them in the habitable zone. However, single systems contain only one planet which transits at small $a$. If it has any companions in the habitable zone, their high $a$ and inclinations yield $|b_\textrm{transit}|>R_{\star}$.

As previously described, \textit{Kepler} is assumed to have 100\% completeness. The \textit{TESS} completeness function which favors small $a$ should further lower the amount of habitable planets observed. However, this effect is too small to be visible in the observed samples, and \textit{TESS} habitability statistics resemble those of \textit{Kepler}.

\section{Discussion}
\label{sec:Discussion}

We have considered the implications of a hypothetical scenario in which dynamical sculpting occurs among M dwarf planetary systems, whether on Myr or Gyr timescales. If the diversity of M dwarf system architectures is, in reality, set in at ``birth", we would have no evidence that the evolutionary timescales of planets in single-transiting systems are different from those in multi-transiting systems. With the assumption of dynamical sculpting on long timescales, the robustness of living organisms to changes in, for example, the planet's orbital eccentricity, are poorly understood. However, we have posited that a ``disruption" event resulting in a planet-planet collision would produce a mass extinction event, after which the processes giving rise to life would need to begin anew. Given this assumption, we consider what we have designated the ``evolutionary timescale" $\tau$, or the duration of time that the planet has enjoyed dynamical quiescence (that is, no orbital changes have recently occurred). The difference in the distributions of evolutionary timescale $\tau$ for singly- and multiply-transiting systems follows from this thought experiment, and may potentially be of future interest. As targets are chosen for atmospheric characterization campaigns, a useful prior to consider, among other factors, is that a planet transiting an M dwarf with no transiting companions is more likely to have had a longer quiescent period than one that does have transiting companions. This is relevant to a sample of M dwarf systems which have their ages uniformly distributed between 0 and 10 Gyr. Ages of M dwarfs are uncertain and difficult to obtain. However, if choosing between planets whose host M dwarfs have ages greater than $\sim4$ Gyr and trying to maximize $\tau$, one would instead favor a dynamically cooler system with multiple transiting companions.  Note that our study assumes that targets will be chosen from the huge catalogs of planets found by the \textit{Kepler} and \textit{TESS} missions, where these findings hold. 

One major assumption in our study is that evolution begins as soon as conditions are conducive to life and is not interrupted by any process other than large-scale dynamical collisions. In reality, evolution will be influenced by factors not considered here such as climate cycles and the cadence of giant impacts \citep{Kopparapu19}. Though the actual process of evolution may be non-linear, we still find it useful to put an upper limit on the amount of this ``uninterrupted'' time a planet would have for potential evolutionary processes. ``Habitability" itself is the subject of necessarily active debate, and whether any of the planets in the \textit{Kepler} and \textit{TESS} samples are, in fact, hosts to living organisms is unknown.

It is important to address the implications of the assumptions made in our study about dynamical sculpting. Our study presents wide constraints on dynamical stability based on observational data. Though the total intact fraction of M dwarf systems is likely between 0.1 and 0.4, it is not yet determined whether systems undergo much dynamical sculpting after formation at all, let alone how quickly whether such sculpting proceeds. Fortunately, our framework to extract estimates for evolutionary timescale can be used with any dynamical sculpting law (i.e., any function of likelihood-of-intactness dependent on age). As more observational data is gathered and theory is advanced, new functions may arise that better reflect reality, and thus more accurately predict evolutionary timescale.

\section{Conclusions}
\label{sec:Conclusions}

Through Monte Carlo simulations of both M dwarf planetary systems and observations of them, we have investigated the hypothetical impact of dynamical sculpting on the potential evolutionary timescale, $\tau$. This quantity encodes the duration of dynamically quiescent time elapsed since a ``disruption" event, and represents an upper limit to the period of time that evolution has proceeded without a collision or other major change to the planet's orbital eccentricity. We found that the rate at which dynamical disruption occurs and the total resulting fraction of systems that are intact at present day significantly affect the distributions of $\tau$ for the entire sample and for different subgroups. Under all aforementioned assumptions, including that the evolutionary clock resets half of the time for planets in disrupted systems, we find that:
\begin{enumerate}
    \item When considering all M dwarf planets observed by \textit{Kepler} or \textit{TESS}, assuming that they are drawn from a uniform age distribution, those transiting in multi systems have average evolutionary timescales $\tau$ lower than or equal to those in single-transiting systems. 
    \item This trend is reversed if an older sample can be identified. Evolutionary timescales $\tau$ of planets transiting in multi systems are greater than or equal to that of planets transiting in single systems, if all systems younger than $\sim$ 4 Gyr are excluded. 
    \item If the rate of dynamical disruption in our galaxy is indeed our ``slow'' rate, and sculpting proceeds over many Gyr, 
    \begin{itemize}
        \item{the average evolutionary timescale $\tau$ of planets in disrupted systems may be as much as $\sim$ 1.4 Gyr greater than that of planets in intact systems.}
        \item{excluding all systems younger than 6 Gyr, the average evolutionary timescale $\tau$ of planets in intact systems may be as much as $\sim$ 0.2 Gyr greater than that of planets in disrupted systems.}
    \end{itemize}
    \item The angular momentum deficit, calculated from relative inclination and eccentricity, of planets in M dwarf systems should increase over time, on average.
\end{enumerate}

We are hopeful that this investigation is a useful contribution to a framework in which orbital excitation, among other myriad properties affecting an exoplanet, is included in considerations of its hospitability to life.

\section{Acknowledgements}
\label{sec:Acknowledgements}
We thank Sarah Rugheimer and Connor Painter for helpful discussions that greatly improved this manuscript. We also thank the University of Florida CLAS Scholars program which funded this project in the 2020-2021 academic year.

\section{Code Availability}
\label{sec:Code Availability}
We make our original code publicly available in a GitHub repository at \url{https://github.com/katieteixeira/evolutionary_timescales}. It can be used to run simulations, save and load data, and make figures.

\section{Appendix}
\label{sec:Appendix}

\subsection{Numerical Predictions}
\label{sec:Numerical-Predictions}
For any LoI function, the distribution of $\tau$ for intact or disrupted systems can be calculated numerically given our assumptions: namely, that (1) our sample of stars is uniformly distributed in age between 0 and 10 Gyr, and (2) that dependent upon the stellar age $A$, a random number $r$, and the collision fraction $f_c$, each system has experienced one of three dynamical outcomes that determine its $\tau$.

For systems that are intact, $\tau$ is always equal to $A$ and, the normalized probability of a $\tau$ is
\begin{equation}
    P(\tau = t~|~S=I) = \frac{\mathscr{L}(t)}{\int_{10^6}^{10^{10}}\mathscr{L}(t)dt}.
\end{equation}

For disrupted systems, the distribution of $\tau$ is complex. It depends on the interplay between $A$, $A_d$,and $f_c$. A given disrupted system has $f_c$ likelihood of being disrupted by ''collision", where $\tau = A - A_d$, and $1-f_c$ likelihood of being disrupted by ``migration", where $\tau = A$. It is useful to visualize the likelihood of these contributing outcomes. Because we choose $A$ and the random value $r$ uniformly, we can visualize Figure \ref{fig:LoI} (albeit with a linear x-axis) as representing all of the possible combinations of $A$ and $r$. Each $(A,r)$ combination determines the dynamical state and $A_d$ (if disrupted), and thus, given collision or migration, determines $\tau$. Given that a system is disrupted, the probability that $\tau$ falls within the finite range ($t_l$, $t_u$) is equal to the fraction of $r$ versus $A$ parameter space that yields $\tau$ in this range.
\begin{equation}
\label{eq:P_tau_in_interval_D}
    P(t_l \leq \tau \leq t_u~|~S=D)= \frac{\mathscr{A}_{cm}+(1-f_c)\mathscr{A}_{m}+f_c\mathscr{A}_{c}}{\int_{10^6}^{10^{10}}\left(1-\mathscr{L}(t)\right)dt}
\end{equation}
The areas $\mathscr{A}$ come from geometric arguments. $\mathscr{A}_{cm}$ is the area of ($A$, $r$) parameter space in which $\tau$ can fall in the interval ($t_l$, $t_u$) due to either disruption mode:

\begin{equation}
    \mathscr{A}_{cm} = \int_{t_l}^{t_u}\Big(1-\mathscr{L}({t-t_l})\Big)dt
\end{equation}

$\mathscr{A}_{m}$ is the area corresponding to systems which have $\tau$ in ($t_l$, $t_u$) solely because they disrupt by migration:
\begin{equation}
    \mathscr{A}_{m} = \int_{t_l}^{t_u}\Big(\mathscr{L}({t-t_l})-\mathscr{L}(t)\Big)dt
\end{equation}
and $\mathscr{A}_{c}$ is the area corresponding to systems which have $\tau$ in ($t_l$, $t_u$) solely because they disrupt by collision:
\begin{equation}
    \mathscr{A}_{c} = \int_{t_u}^{10^{10}}\Big( \mathscr{L}\big({t-t_u}\big)-\mathscr{L}(t-t_l)\Big) dt
\end{equation}

\begin{figure*}
    \centering
    \includegraphics[trim={0.8cm 0 0.8cm 0},width=4in]{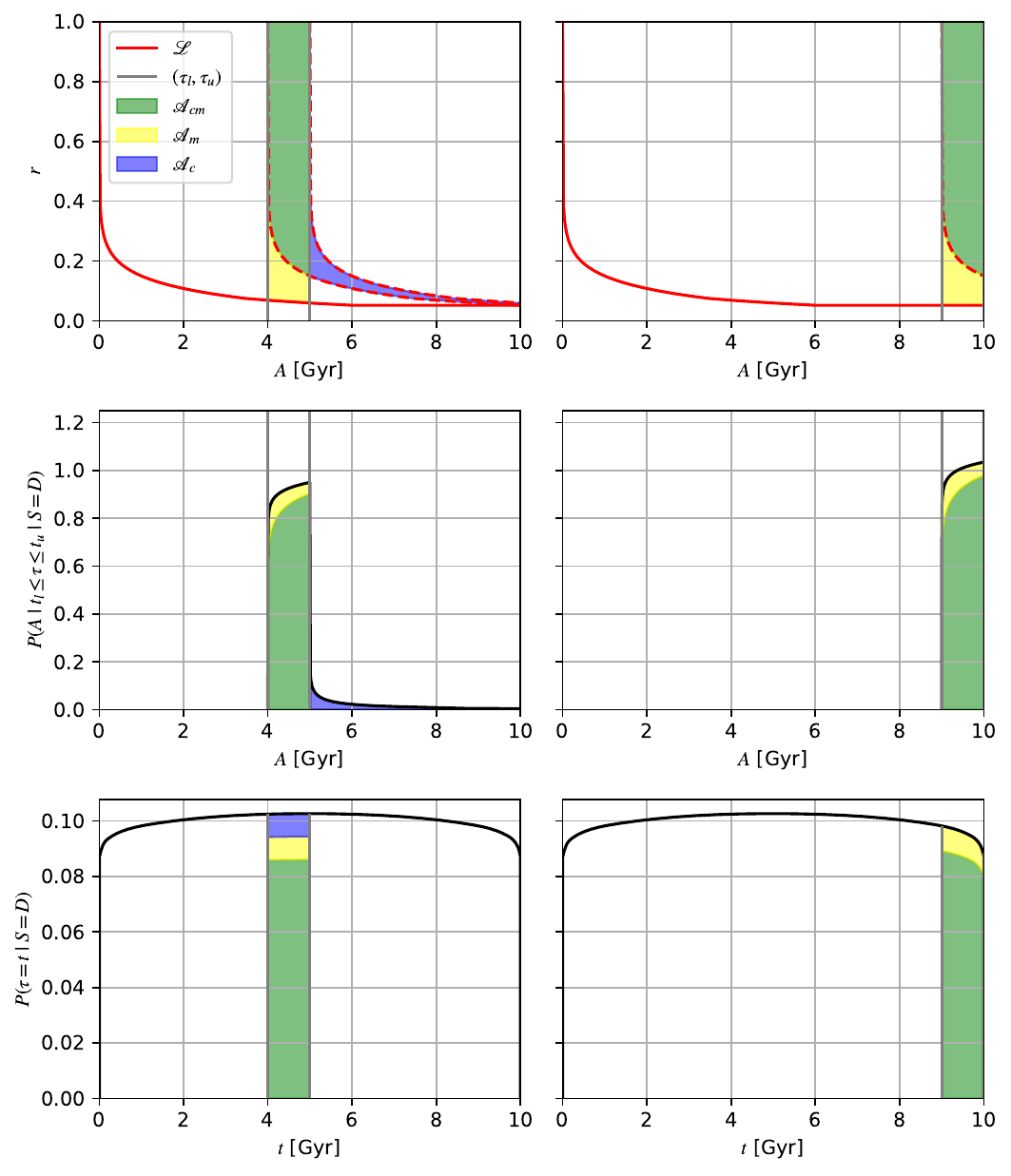}
    \caption{Figure showing how two different ranges of $\tau$ have different probabilities in a representative sample where $\bar{\mathscr{L}}=0.1$, $\mathscr{D}=$ slow, and $f_c=0.5$. The left panel shows $t_l=4~\mathrm{Gyr}$, $t_u=5~\mathrm{Gyr}$ and the right shows $t_l=9~\mathrm{Gyr}$, $t_u=10~\mathrm{Gyr}$. \textit{Top}: Plot of $r$ versus $A$ parameter space showing which portions yield $\tau$ values in the range. $(A,r)$ values that fall within $\mathscr{A}_{cm}$ are certain to yield $\tau$ in this range. However, systems with $(A,r)$ in $\mathscr{A}_m$ or $\mathscr{A}_c$ must disrupt by migration or collision, respectively, to be included. The total probability that a system has $\tau$ in this range is a function of each of these areas, weighted by $f_c$. \textit{Middle}: The normalized probability of a certain age given that $\tau$ is in the range and the system is disrupted, highlighting the range's respective contributions from the different areas $A$. \textit{Bottom}: The numerically calculated distribution of $\tau$ for disrupted systems, again highlighting the range and its respective contributions from the different areas $A$.}
    \label{fig:Areas-Explanation}
\end{figure*}

These areas in $r$ versus $A$ parameter space are displayed in the first row of Figure \ref{fig:Areas-Explanation} for $t_l=4~\mathrm{Gyr}$, $t_u=5~\mathrm{Gyr}$ and $t_l=9~\mathrm{Gyr}$, $t_u=10~\mathrm{Gyr}$ where $\bar{\mathscr{L}}=0.1$ and $\mathscr{D}=$ slow. A combination of ($A$, $r$) which lands in the $\mathscr{A}_{cm}$ space will result in $\tau$ in this range regardless of disruption mode. A combination which lands in $\mathscr{A}_{m}$ will result in $\tau$ in this range only if the disruption mode is migration, and that which lands in $\mathscr{A}_{c}$ will result in $\tau$ in this range only if the disruption mode is collision. The second row of Figure \ref{fig:Areas-Explanation} shows the normalized probability of a certain age given that $\tau$ is in the range ($t_l$, $t_u$) and the system is disrupted, assuming $f_c=0.5$.

Calculating the distribution $P(\tau=t~|~S=D)$ requires computing \eqref{eq:P_tau_in_interval_D} for many small intervals ($t-dt$, $t+dt$) subdividing $10^6$ to $10^{10}$ years, then normalizing. The bottom row of Figure \ref{fig:Areas-Explanation} shows this for $\bar{\mathscr{L}}=0.1$, $\mathscr{D}=$ slow, and $f_c=0.5$, still highlighting the two different ranges ($t_l$, $t_u$) and their respective contributions from the different areas $A$. Because the $\tau$ range between 9 and 10 Gyr has no contribution from $\mathscr{A}_c$, we see that the resulting value of $P(\tau=t~|~S=D)$ is lower here than for a range between 4 and 5 Gyr.

\begin{figure*}
    \centering
    \includegraphics[width=7.0in]{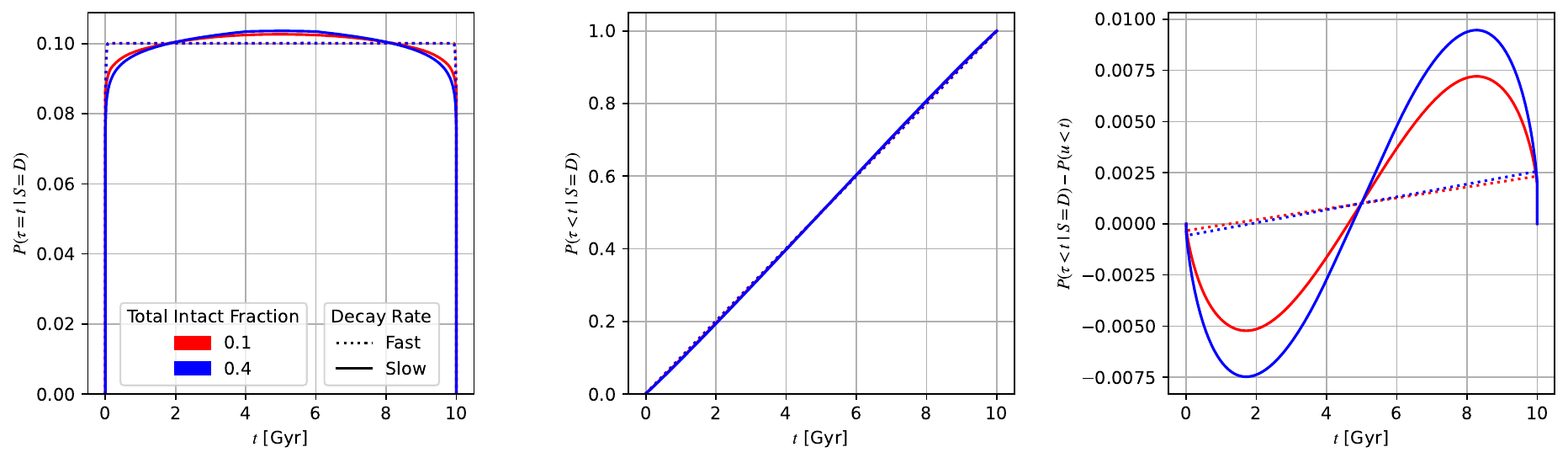}
    \caption{Numerically calculated distributions of $\tau$ for disrupted systems where $f_c=0.5$, displayed as a probability density function (left), a cumulative density function (middle), and the difference between the cumulative density function and that of a uniform distribution (right).}
    \label{fig:Numerical-Prediction}
\end{figure*}

The distributions $P(\tau=t~|~S=D)$ for each of our limiting combinations of $\bar{\mathscr{L}}$ and $\mathscr{D}$ are shown in Figure \ref{fig:Numerical-Prediction} along with the cumulative distribution and the difference between the cumulative distribution and a uniform one. Each exhibit a peak close to 5 Gyr with lower probability and higher $\tau$. This is because the sum of $\mathscr{A}_m$ and $\mathscr{A}_c$ reaches its maximum at this intermediate $\tau$ value. At the smallest $\tau$, $\mathscr{A}_m=0$, and at the largest $\tau$, $\mathscr{A}_c=0$, as explained previously and visualized in Figure \ref{fig:Areas-Explanation}. The effect exists for each $\bar{\mathscr{L}}$, $\mathscr{D}$ combination but is least visible for $\mathscr{D}$=fast which is almost a uniform distribution.

\subsection{Additional Figures}
\begin{figure*}
    \centering
    \includegraphics[width=7.0in]{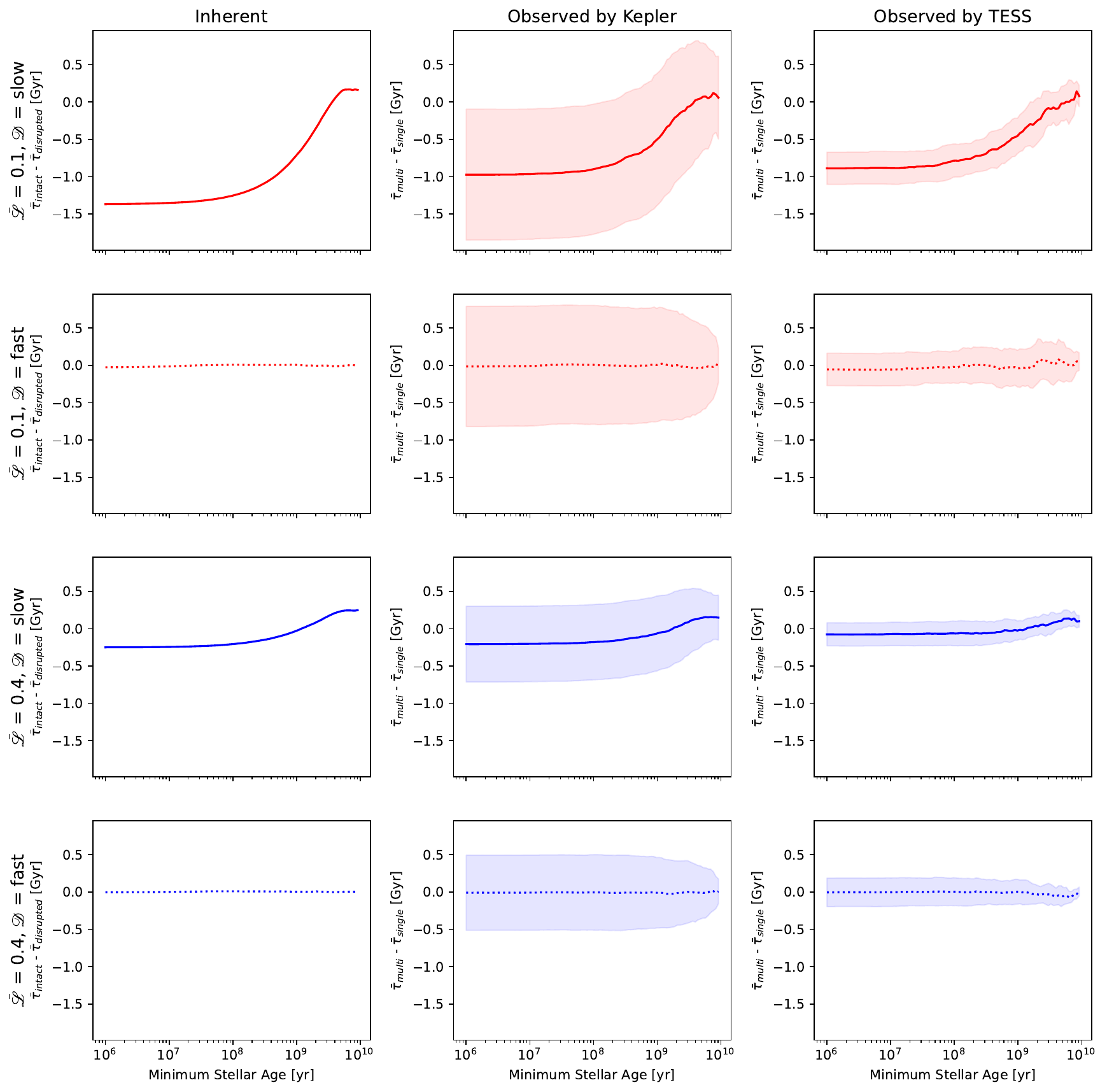}
    \caption{Same as Figure \ref{fig:Delta-Tau-Vs-Min-Age} but with error shown. The $+/-$ error shown here is the quadrature of the standard deviation of intact/multi $\tau$ and that of disrupted/single $\tau$ over the collection of simulations.}
    \label{fig:Delta-Tau-Vs-Min-Age-w-Error}
\end{figure*}

We provide one additional figure in this appendix, Figure \ref{fig:Delta-Tau-Vs-Min-Age-w-Error}, which is identical to Figure \ref{fig:Delta-Tau-Vs-Min-Age}, with error bars.

\bibliography{main.bib}

\end{document}